\documentclass[aps,twocolumn,nofootinbib,prd]{revtex4}
\usepackage{graphicx}
\usepackage{amsfonts}
\usepackage{amssymb}
\usepackage{amsbsy}
\usepackage{amsmath}
\usepackage{latexsym}
\usepackage{natbib}
\usepackage{bm}
\usepackage{accents}
\usepackage{color}
\usepackage{hyperref}
\usepackage{float}

\begin{document}
\title{A wormhole geometry from gravitational collapse}

\author{Soumya Chakrabarti\footnote{soumya.chakrabarti@saha.ac.in}}
\affiliation{Theory Division,\\
Saha Institute of Nuclear Physics, \\
Kolkata 700064, West Bengal, India
}

\author{Sayan Kar\footnote{sayan@phy.iitkgp.ac.in}}
\affiliation{Department of Physics, \\
Indian Institute of Technology Kharagpur, 721302, India.}

\date{\today}

\begin{abstract}
We discuss a proposal on how gravitational collapse of a NEC (Null Energy Condition) violating spherically symmetric fluid distribution can avoid the formation of a zero proper volume singularity and eventually lead to a Lorentzian wormhole geometry. Our idea is illustrated using a time-evolving wormhole spacetime in which, we show how a collapsing sphere may never reach a zero proper volume end-state. The nature of geodesic congruences in such spacetimes is considered and analysed. Our construction is inspired from a recently proposed static wormhole geometry, the multi-parameter Simpson-Visser line element, which is known to unite wormholes and black holes (regular and singular) in a single framework.
\end{abstract}

\maketitle

\section{Introduction}
Gravity provides a possibility of collapse without limit due to its attractive nature. For a sufficiently massive distribution, gravitational collapse can happen when gravity starts to dominate over the internal pressure. What forms at the end of a collapse, has a long-standing issue of predictability and remains an unresolved question even in the current context. In the $1930$s, Chandrasekhar \cite{chandra} and Landau \cite{landau} suggested an upper limit to the mass of a collapsing object beyond which it cannot further support itself against the inward pull of gravity. Landau \cite{landau} and Eddington \cite{eddi} showed that a massive stellar distribution exceeding the Chandrasekhar limit can shrink indefinitely until it forms an end state from which even light cannot escape. The first complete study of a star undergoing gravitational collapse to zero proper volume within the framework of General Theory of Relativity (GR) was given by Datt \cite{datt}, Oppenheimer and Volkoff \cite{oppenvol}, Oppenheimer and Snyder \cite{oppensny} and their works laid the foundation of the physics of gravitational collapse. A spacetime singularity is predicted to form at the end of gravitational collapse where physical quantities like density, pressure as well as geometrical quantities like the curvature scalars blow up. In general, such a collapsed region is expected to remain enveloped within a horizon barring all information exchanges with a faraway observer. Such an end state is known as a black hole. It remains unclear if under increasingly generalized conditions a massive star would necessarily end up in a black hole or not. A novel attempt at solving this issue came in the form of `Cosmic Censorship Hypothesis' \cite{penrose}, according to which, the ultimate spacetime singularity is necessarily accompanied with the formation of a horizon. However, there exists no proof of the conjecture and a number of counterexamples suggest that, in principle, the singularity can also remain exposed to a faraway observer. Such a concept is called a `naked singularity' (Yodzis, Seifert et. al. \cite{yod}; Eardley and Smarr \cite{eard}; Christodoulou \cite{christ}; Joshi et. al. \cite{joshi}; Waugh and Lake \cite{waugh}; Ori and Piran \cite{ori}; Joshi et. al. \cite{joshietal}). The problem as of today stands without any complete theorem that can clearly pick one out of these two possibilities as the possible end-state.   \\

However, besides the above, other possibilities are also considered through specific examples. For instance, Choptuik's study on the collapse of a scalar field minimally coupled to gravity led to the discovery of `critical phenomena in gravitational collapse' \cite{chop} (in analogy with critical phenomena in phase transitions) where the outcome depends on the value of one or more parameters defining the initial condition. For some values of the parameter there is no spacetime singularity as the scalar field collapses and eventually disperses leaving behind a flat space \cite{critical}. Higher curvature corrections to GR can also produce some interesting examples, for example, in an $f(R)$ theory, collapsing solutions in general mimic it's GR counterpart, for example a dust collapse in $f(R)$ gravity mimics (\cite{ritu}) the Lemaitre-Tolman-Bondi \cite{ltb} solution of GR. However, it has been proved recently that conformal symmetry can drive a collapsing star in $f(R)$ gravity with anisotropic pressure and heat flux towards a non-singular end-state \cite{scfr}. A Gauss-Bonnet invariant non-minimally coupled to GR can also lead to interesting signatures, an area which is gaining increasing attention quite recently, in studies of a singularity-free model of gravitational collapse \cite{scsegb}.    \\

In the present manuscript, we propose another interesting possibility - a gravitational collapse ending with a nonsingular wormhole geometry. We support this claim with an example of a non-static `solution' of the Einstein field equations which evolves into a static wormhole. A wormhole is a concept that defines pathways connecting distant parts of the same universe or two different universes, through a `throat' \cite{morris1, visser1}. The geometric structure is tubular which spreads out and becomes asymptotically flat at the infinities. Converging radial null rays become parallel at the wormhole throat and eventually diverge on the other side. In GR, the Schwarzschild solution is a wormhole as was hinted in \cite{flamm}, although the throat of the Schwarzschild wormhole pinches off leading to a singularity. In their seminal work, Einstein and Rosen, while trying to build a particle model in GR, used the wormhole idea for the first time. Thus, the Einstein-Rosen bridge \cite{einsteinrosen} is cited as an early precursor of the modern-day wormhole. The term `wormhole' was however introduced only in the late $1950$-s in the work on geons and a topological model of electric charge {\em a la} Wheeler \cite{misnerwheeler}. In a wormhole geometry, radial null rays defocus at and around the throat, essentially due to the violation of Null Convergence Condition (NCC) \cite{visser1, misnerwheeler}. Using GR one can realize that a violation of the NCC leads to a violation of the Null Energy Condition (NEC) \cite{hawking, kolassis, pimentel}). Most examples of wormhole solutions in  the literature abide by this violation. For instance, the Ellis-Bronnikov wormhole exists for a phantom (negative kinetic energy) scalar field \cite{ellis, bronnikov}. However, there are models where the violation of NCC does not necessarily imply an energy condition violation \cite{capo}. Bhawal and Kar \cite{sk1}, Maeda and Nozawa \cite{maeda} studied the existence of a Lorentzian wormhole in Einstein-Gauss-Bonnet theory which does not violate NEC. Kar, Lahiri and Sengupta \cite{sk2} considered the role of extra-dimensional effects on wormhole solutions without any exotic matter. The concept of a wormhole also leads to time machine models which further enhance interest in the subject \cite{morris2, novikov, frolov}. More such examples and different aspects of wormhole geometries have already been discussed and summarized in recent literature, quite extensively \cite{literature}.   \\

We focus on the recently proposed Simpson-Visser solution \cite{simpsonvisser}. This is a modification of the standard Schwarzschild spacetime, with an additional parameter being introduced in the metric which controls the interpolation of the metric between a standard Schwarzschild black hole and a Morris-Thorne traversable wormhole. Our primary aim is to look into the time evolution of an initially collapsing geometry that ends up in a similar geometry. We primarily look into the behavior of geodesic congruences and confirm that no zero proper volume singularity is reached at any time. Moreover, from a suitable boundary matching condition we give an exact collapsing solution, which slowly evolves into a spherical wormhole geometry at a non-zero minimum radius. We discuss that this singularity-free nature of the spacetime lies within its wormhole-like structure, which also leads to a violation of the Null Convergence Condition. Similar scenarios were also constructed in a rather old paper by Roman and Bergmann \cite{roman} who gave a model of spherically symmetric collapse without any singularity as well as a violation of the weak energy condition. In a sense, this present manuscript can serve as an extension of such an analysis as we show that the inhomogeneity of the collapsing core can have different forms, keeping the qualitative outcome of the collapse similar, i.e., non-singular. \\

Our article is organised as follows. In Section II, we recall the basics on the Simpson-Visser line element and introduce the time evolving generalizations. Section III discusses the geodesic congruences in such spacetimes and what we can infer from them. An exact solution is discussed along with the energy conditions in Section IV. Finally, we conclude with some remarks in Section V. 

\section{The Simpson-Visser spacetime}
Our basic mathematical setup begins with the notion of a regular black hole \cite{Bardeen:1968, Bergmann-Roman, Hayward:2005}. Regular black holes are non-singular in nature. We concentrate on one particular candidate, the solution of Simpson and Visser \cite{simpsonvisser} given by the metric

\begin{eqnarray}\nonumber \label{simpsonvisser}
&&ds^{2} = \Big(1-\frac{2m}{\sqrt{l^{2}+a^{2}}}\Big)dt^{2} - \frac{dl^{2}}{1-\frac{2m}{\sqrt{l^{2}+a^{2}}}} \\&& - \Big(l^{2}+a^{2}\Big)\Big(d\theta^{2}+\sin^{2}\theta d\phi^{2}\Big).
\end{eqnarray}

In this definition, the coordinates can vary over the domains
\begin{eqnarray}\nonumber
&&l\in(-\infty,+\infty);\\&&\nonumber 
t\in(-\infty,+\infty).
\end{eqnarray}

For different values of the parameter $a$, the metric can yield different geometric structures. When $a = 0$ we get the standard Schwarzschild metric. To study either a regular black hole or a traversable wormhole $a\neq 0$ is the primary condition. Radial null rays in the metric can be obtained from,
\begin{equation}
\frac{dl}{dt} = \pm\Big(1-\frac{2m}{\sqrt{l^{2}+a^{2}}}\Big).
\end{equation}

\begin{itemize}
\item {For $a > 2m$, all $l \in (-\infty, +\infty)$ gives $\frac{dl}{dt}\neq 0$. Therefore the geometry indicates a traversable wormhole where the wormhole throat is a timelike hypersurface at $l=0$. $l < 0$ corresponds to a universe on the other side of the geometry.}
\item {For $a = 2m$, $l\rightarrow 0$ always implies $\frac{dl}{dt}\rightarrow 0$ for both positive and negative $l$. Therefore, a horizon exists at $l = 0$ indicating the extremal null throat of a one-way wormhole.}
\item {For $a < 2m$, $\frac{dl}{dt} = 0$ for the values of $l$ satisfying $\sqrt{l_\pm^{2} + a^{2}} = 2m$. The $l$-coordinate values corresponding to a pair of horizons are given by $l_{\pm} =\pm\sqrt{(2m)^{2}-a^{2}}$.}
\end{itemize}

There is no curvature singularity. At $l = 0$ we obtain a finite value of the curvature tensor components and the scalar invariants.  \\

Let us now rewrite the metric in a slightly different set of coordinates. A coordinate transformation given as
\begin{equation}
l^{2} + a^{2} = r^{2},
\end{equation}
transforms the metric in Eq. (\ref{simpsonvisser}) into 
\begin{equation}\label{beginning}
ds^{2} = \Big(1-\frac{2m}{r}\Big)dt^{2} - \frac{dr^{2}}{\Big(1 - \frac{a^{2}}{r^{2}}\Big)\Big(1-\frac{2m}{r}\Big)} - r^{2} d\Omega^2.
\end{equation}
The radial coordinate can now vary over the domain defined below
\begin{eqnarray}
r\in(a,+\infty).
\end{eqnarray}
The solution in Eq. (\ref{beginning}) is a static solution. A time-evolving metric extension of this geometry may be taken as, 
\begin{equation} \label{beginning2}
ds^{2} = A(r,t)^{2} dt^{2} - \frac{dr^{2}}{\Big(1 - \frac{b(r)}{r}\Big)B(r,t)^2} - C(r,t)^{2} d\Omega^2,
\end{equation}
where the coordinates can vary over the domains
\begin{eqnarray}\label{coordinate}\nonumber
&&r\in(r_{w},+\infty);\\&& 
t\in(0,+\infty),
\end{eqnarray}

where $r_{w}$ is supposed to be the throat of the wormhole, equivalent to $a$. Our intention here is to study the evolution of an initially collapsing sphere having a geometry similar to a simpified form of Eq. (\ref{beginning2}) and see if it can lead to a geometry similar to Eq. (\ref{beginning}) as the end-state of collapse. 

\section{Evolution of Geodesic Congruences}
The final fate of a collapse can be studied using kinematic variables such as expansion, shear and rotation of a geodesic congruence in the geometry. Evolution of such quantities along a geodesic congruence are governed by the Raychaudhuri equations \cite{rc}. Although the kinematic quantities for geodesic congruences are well-studied for static spacetime geometries \cite{congrusk}, a non-static collapsing spacetime has received less attention in this regard \cite{rajibulsk}. 

\noindent Let us begin with the expansion, defined as the rate of change of the cross-sectional area orthogonal to the geodesic flow. Assuming the line element as
\begin{equation}\label{metricgeodesic1}
ds^{2} = T(t)^{2} \Bigg[A(r)^{2}dt^{2} - B(r)^{2}dr^{2} - r^{2}d\Omega^2 \Bigg],
\end{equation}
we note that we have considered the metric coefficients as separable functions of $r$ and $t$, for the sake of simplicity. Moreover, the time evolution of the metric comes in only as a conformal factor. Spacetime geometries with a similar symmetry have been found in studies on radiating fluid spheres, shear-free stars and scalar field collapse \cite{metricchoice}. Here, to realize the underlying wormhole structure one can replace $B(r)$ with $\frac{1}{\big(1 - \frac{b(r)}{r}\big)^{\frac{1}{2}}A(r)}$. The velocity vector is $u^\alpha=(t^{\circ},r^{\circ},\phi^{\circ})$ where `$\circ$' is the derivative with respect to an affine parameter $\tau$. It satisfies the constraint 
\begin{equation}
u^\alpha u_\alpha=s, 
\end{equation}
where $s = 1$ is for timelike geodesics and $s = 0$ is for null geodesics. Assuming $\theta = \frac{\pi}{2}$, we find that the geodesics in the collapsing spacetime have to obey the following equations.
\begin{equation}
\phi^{\circ} = \frac{L}{r^{2}T^{2}},
\label{eq:geodesic_phi}
\end{equation}
\begin{equation}
r^{\circ} = \left\lbrace \frac{A^2}{B^2}{t^{\circ}}^{2} - \frac{(s r^{2}T^{2} + L^{2})}{r^{2}B^{2}T^{4}} \right\rbrace^{\frac{1}{2}},
\label{eq:geodesic_r}
\end{equation}
and
\begin{equation}
t^{\circ\circ} + 2\frac{\dot{T}}{T}{t^{\circ}}^{2} + 2\frac{A'}{A}{t^{\circ}}{r^{\circ}} - \frac{s \dot{T}}{A^{2}T^{3}} = 0,
\label{eq:geodesic_t}
\end{equation}
where a `prime' and a `dot' denote derivatives with respect to $r$ and $t$, respectively. $L$ is a constant of integration. The kinematic quantities are in general studied on a spacelike hypersurface orthogonal to the congruence, using an induced metric $h_{\alpha\beta}$ defined as
\begin{equation}
h_{\alpha\beta}=g_{\alpha\beta}-u_\alpha u_\beta, \hspace{1.0cm} (\alpha,\beta=0,1,2,3).
\label{eq:transverse_metric}
\end{equation}
where $u^\alpha$ is timelike and tangent to points on each geodesic. The transverse metric is orthogonal to $u^\alpha$. We define the velocity gradient tensor as
\begin{equation}
B_{\alpha\beta}=\nabla_{\beta} u_\alpha.
\label{eq:vel_grad}
\end{equation}
The tensor $B_{\alpha\beta}$ can be expressed as a sum of three parts, trace, symmetric traceless and antisymmetric and be written as
\begin{equation}
B_{\alpha\beta}=\frac{1}{3}h_{\alpha\beta}\Theta+\sigma_{\alpha\beta}+\omega_{\alpha\beta},
\label{eq:bdef}
\end{equation}
where
\begin{eqnarray}\label{thetastart}
&&\Theta = B^\alpha_{\;\alpha}, \\&&
\sigma_{\alpha\beta}=\frac{1}{2}(B_{\alpha\beta}+B_{\beta\alpha})-\frac{1}{3}h_{\alpha\beta}\theta, \\&&
\omega_{\alpha\beta}=\frac{1}{2}(B_{\alpha\beta}-B_{\beta\alpha}).
\end{eqnarray}
$\Theta$ is the expansion scalar, $\sigma_{\alpha\beta}$ is the shear tensor and $\omega_{\alpha\beta}$ is the rotation tensor. By virtue of this construction, the shear and the rotation tensors satisfy $h^{\alpha\beta}\sigma_{\alpha\beta} = 0$ and $h^{\alpha\beta}\omega_{\alpha\beta} = 0$. We also have $g^{\alpha\beta}\sigma_{\alpha\beta} = 0$ and $g^{\alpha\beta}\omega_{\alpha\beta} = 0$. The evolution equation for the spatial tensor $B_{\alpha\beta}$ can be written as,
\begin{equation}
u^\gamma \nabla_\gamma B_{\alpha\beta}=-B_{\alpha\gamma}B^\gamma_{\;\beta}+R_{\gamma\beta\alpha\delta}u^\gamma u^\delta,
\label{eq:evolution_eqn}
\end{equation}
where $R_{\gamma\beta\alpha\delta}$ is the Riemann tensor. We are particularly interested in the trace part of the above equation which governs the evolution of $\Theta$,
\begin{equation}
\frac{d\Theta}{d\tau} + \frac{1}{3}\Theta^{2} + \sigma^2 - \omega^2 + R_{\alpha\beta}u^\alpha u^\beta = 0.
\label{eq:expansion_eqn}
\end{equation}
$R_{\alpha\beta}$ is the Ricci tensor, $\sigma^2 = \sigma^{\alpha\beta}\sigma_{\alpha\beta}$ and $\omega^2=\omega^{\alpha\beta}\omega_{\alpha\beta}$. The equation for the expansion can predict the tendency of a family of initially converging geodesics to focus within a finite `time'--a result known as the focusing theorem \cite{rc}. An expansion approaching negative infinity could imply a mere convergence of the bundle or a focusing towards a genuine singularity. Using the metric in Eq. (\ref{metricgeodesic1}) and the geodesic Eqs. (\ref{eq:geodesic_phi}), (\ref{eq:geodesic_r}) and (\ref{eq:geodesic_t}) we write the expansion for the geodesic congruence as,
\begin{equation}\label{thetageneral1}
\Theta = u^{\alpha}_{;\alpha} = \frac{s\dot{T}}{A^{2}T^{3}t^{\circ}} + A(r) \Big(1 - \frac{b(r)}{r} \Big)^{\frac{1}{2}} W(r,t),
\end{equation}
where
\begin{eqnarray}\label{thetageneral2}\nonumber
&& W(r,t) = \Bigg\lbrace \Bigg(2A^{2}{t^{\circ}}^{2} - \frac{L^{2}}{r^{2}T^{4}}\Bigg) \Bigg(\frac{1}{r} - \frac{A'}{A}\Bigg) + \frac{s}{T^{2}}\\&&
\Bigg(\frac{A'}{A} - \frac{2}{r}\Bigg) \Bigg\rbrace \Bigg\lbrace A^{2}{t^{\circ}}^{2} - \frac{L^{2}}{r^{2}T^{4}} \Bigg(1 + \frac{s r^{2}T^{2}}{L^2}\Bigg) \Bigg\rbrace^{-\frac{1}{2}}.
\end{eqnarray}

\begin{enumerate}
\item {From the rather complicated expression above, we note that since $\Big\lbrace A^{2}{t^{\circ}}^{2} - \frac{L^{2}}{r^{2}T^{4}} \Big(1 + \frac{s r^{2}T^{2}}{L^2}\Big) \Big\rbrace^{-\frac{1}{2}}$ has to be real, we require 
\begin{equation}
\Bigg\lbrace A^{2}{t^{\circ}}^{2} - \frac{L^{2}}{r^{2}T^{4}} \Big(1 + \frac{s r^{2}T^{2}}{L^2}\Big) \Bigg\rbrace \geq 0.
\end{equation}
The above condition translates into
\begin{equation}
r^{2}A^{2}{t^{\circ}}^{2}T^{4} - sr^{2}T^{2} - L^{2} \geq 0.
\end{equation}
Using the equality sign in the above equation and the standard way of writing roots of a quadratic equation, we can constrain the minima of the time evolving conformal factor as 
follows:
\begin{equation}
T^{2} \geq \frac{s}{2A^{2}{t^{\circ}}^{2}} \pm \frac{\Big(r^{2}s^{2} + 4A^{2}{t^{\circ}}^{2}L^{2}\Big)^{\frac{1}{2}}}{2rA^{2}{t^{\circ}}^{2}}. 
\end{equation}
}
\item {For a static wormhole with $A(r) = 1$, from Eq. (\ref{eq:geodesic_t}) we note that $t^{\circ} = \alpha$, where $\alpha$ is a constant. Therefore, the expansion for a static metric becomes
\begin{equation}
\Theta = \frac{2(\alpha^{2} - s)^{\frac{1}{2}}}{r} \Big(1 - \frac{b(r)}{r} \Big)^{\frac{1}{2}}.
\end{equation}
This vanishes at the throat $r = b(r)$ and goes to zero asymptotically.
}
\item {Overall, the first term on the RHS of Eq. (\ref{thetageneral1}), $\frac{s\dot{T}}{A^{2}T^{3}t^{\circ}}$, is negative for an initially collapsing geometry. While for a homogeneous spacetime this definitely drives $\Theta$ towards negative $\infty$, indicating a formation of singularity, in the present case, due to the inhomogeneity in $g_{00}$ and $g_{11}$, it is quite possible for the second term on the RHS $\frac{1}{B(r)} W(r,t)$ to be positive and dominating. As a result the formation of a singularity can be avoided.}
\end{enumerate} 

The special case of $T(t) = 1$, i.e., the static case, deserves a brief discussion. In this case, retaining the general form of $A(r)$ and $B(r)$, one essentially gets a static Simpson-Visser metric, resembling the end-geometry of the non-singular collapse. We note from Eq. (\ref{eq:geodesic_t}) that $t^{\circ} = \frac{\alpha}{A^{2}}$, where $\alpha$ is a constant. Therefore the expansion for this static line element becomes
\begin{equation}\label{staticexp}
\Theta = A(r) \Big(1 - \frac{b(r)}{r} \Big)^{\frac{1}{2}} W_{1}(r),
\end{equation}
where
\begin{eqnarray}\nonumber
&& W_{1}(r) = \Bigg\lbrace \Bigg(2\frac{\alpha^2}{A^2} - \frac{L^{2}}{r^{2}}\Bigg) \Bigg(\frac{1}{r} - \frac{A'}{A}\Bigg) + s\Bigg(\frac{A'}{A} - \frac{2}{r}\Bigg) \Bigg\rbrace \\&&
\Bigg\lbrace \frac{\alpha^2}{A^2} - \frac{L^{2}}{r^{2}} \Bigg(1 + \frac{s r^{2}}{L^2}\Bigg) \Bigg\rbrace^{-\frac{1}{2}}.
\end{eqnarray}
For a real expression of $\Theta$ we must ensure that
\begin{equation}
\Bigg\lbrace \frac{\alpha^2}{A^2} - \frac{L^{2}}{r^{2}} \Bigg(1 + \frac{s r^{2}}{L^2}\Bigg) \Bigg\rbrace^{-\frac{1}{2}} \geq 0.
\end{equation}
Using the equality sign in the above equation and the standard way of writing roots of a quadratic equation, we constrain metric components as
\begin{equation}
\alpha^{2}r^{2} - L^{2}A^{2} - sA^{2}r^{2} \geq 0,
\end{equation}
which essentially means that $r$ can not reach $0$, since $- L^{2}A^{2}$ can not be positive. The minimum allowed radial value of the sphere is
\begin{equation}
r \geq \left( \frac{L^2 A^2}{\alpha^{2} - sA^{2}}\right)^{1/2}.
\end{equation}

Using the functional form of $A(r)$ from Eq. (\ref{beginning}), we simplify the static expansion as in Eq. (\ref{staticexp}) to plot as a function of radial coordinate $r$.

\begin{figure}[h]
	\begin{center}
		\includegraphics[angle=0, width=0.35\textwidth]{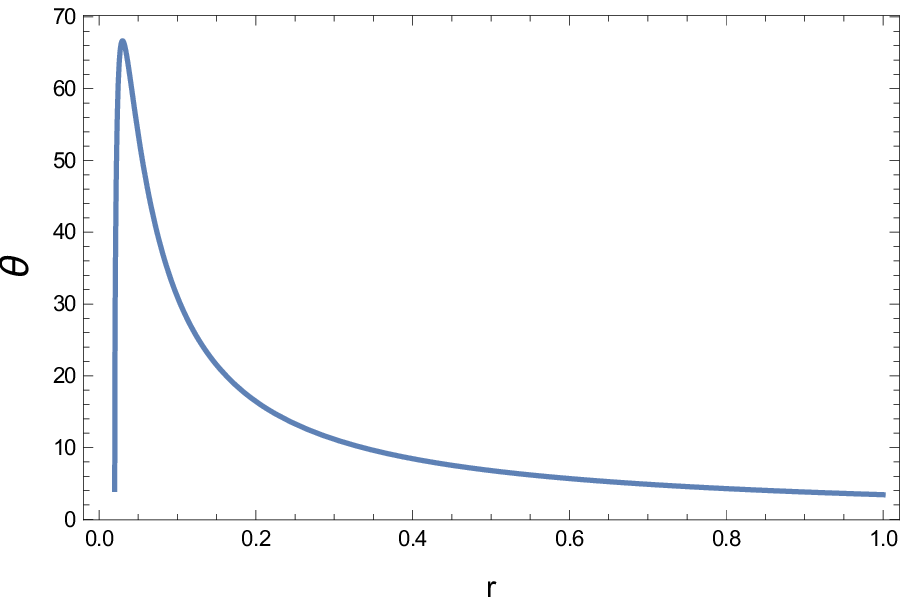}
		\includegraphics[angle=0, width=0.35\textwidth]{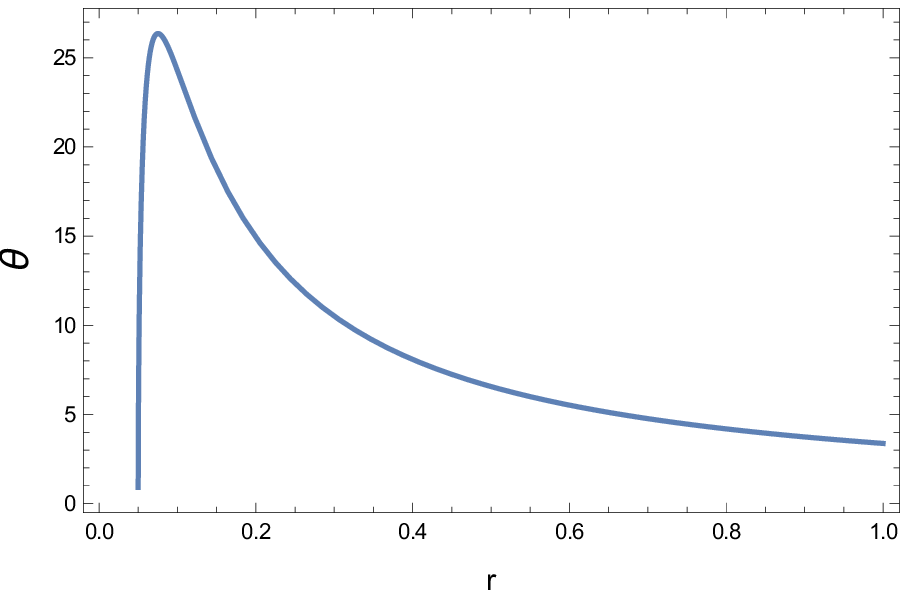}
		\includegraphics[angle=0, width=0.37\textwidth]{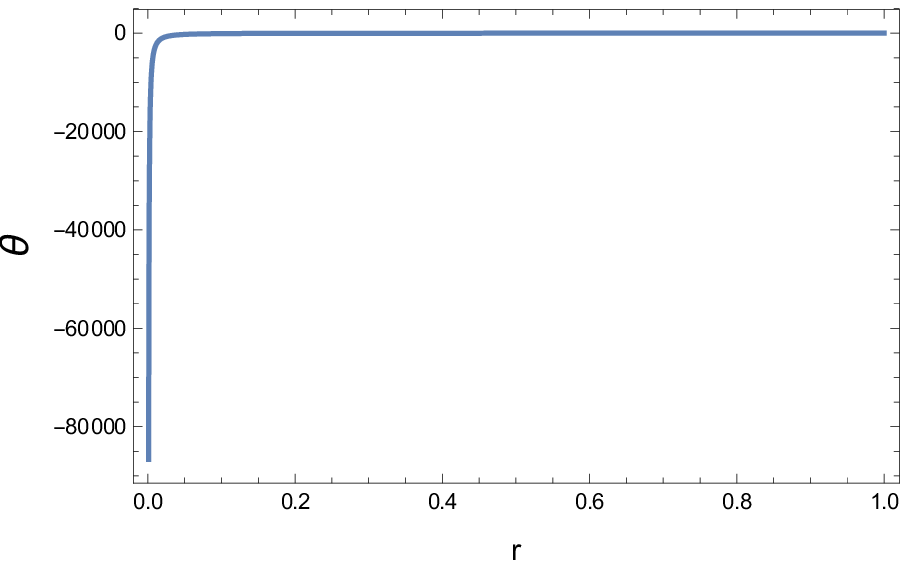}
	\end{center}
	\caption{{\small Evolution of $\Theta(r)$ as a function of $r$. The three graphs signify evolution for $a \neq 0$, $m = 0$ (top), $a > m \neq 0$ (middle) and $a = 0$ (bottom). For the graph on top $a = 0.03$ and for the graph in middle, $a = 0.05$, signifying the wormhole throats for the static Ellis wormhole and Simpson-Visser Wormhole solutions, respectively.}}
	\label{theta_static}
\end{figure}

In Fig. \ref{theta_static}, we plot the radial distribution of the expansion (for the time-independent spacetimes) in the range $0 \leq r \leq 1$. The graph on top shows the expansion where we have chosen $m = 0$, which signifies an Ellis wormhole. The graph in the middle is for $a > m$ and both of these parameters are non-zero. This essentially is a static Simpson-Visser non-singular geometry. We can see that $\Theta$ never reaches $-\infty$ for these two cases and around $r = a$, the wormhole throat forms. The graph at the bottom is for $a = 0$, which essentially switches off the characteristics of a wormhole geometry and leaves behind a Schwarzschild Black Hole. Therefore it gives us $\Theta \rightarrow -\infty$ around $r \sim 0$. \\
 
From Eq. (\ref{eq:expansion_eqn}), with $\sigma^2 \geq 0$ for a congruence of timelike curves, we recall the timelike convergence condition given as,
\begin{equation}\label{ncc}
R_{\alpha\beta}u^\alpha u^\beta \geq 0.
\end{equation}
This is required for geodesic focusing to happen. In the present setup this general condition is written as
\begin{eqnarray}\nonumber\label{nccdetail}
&& R_{\alpha\beta}u^\alpha u^\beta = - \frac{d\Theta}{d\tau} - \frac{1}{3}\Theta^{2} - \sigma^2 + \omega^2 \\&&\nonumber
= \Bigg[3\Bigg(\frac{\dot{T}^{2}}{T^2} - \frac{\ddot{T}}{T}\Bigg) + \frac{A}{rB^3} \lbrace -rA'B' + B(2A' + rA'') \rbrace \Bigg] \\&&\nonumber
{t^{\circ}}^{2} + \Bigg[\frac{B^{2}}{A^{2}}\Bigg(\frac{\dot{T}^{2}}{T^2} + \frac{\ddot{T}}{T}\Bigg) + \frac{1}{AB}(A'B'-A''B) + \frac{2B'}{rB}\Bigg] \\&&\nonumber
\Bigg\lbrace \frac{A^2}{B^2}{t^{\circ}}^{2} - \frac{(L^{2} + r^{2}T^{2})}{r^{2}B^{2}T^{4}} \Bigg\rbrace + \Bigg[1 + \frac{r^{2}}{A^{2}}\Bigg(\frac{\dot{T}^{2}}{T^2} + \frac{\ddot{T}}{T}\Big) \\&&\nonumber
+ \frac{rB'}{B^3} - \frac{(A + rA')}{AB^2}\Bigg] \frac{L^{2}}{r^{4}T^{4}} + 4\frac{A'\dot{T}}{AT}t^{\circ}\Bigg\lbrace \frac{A^2}{B^2}{t^{\circ}}^{2} \\ &&
 - \frac{(L^{2} + r^{2}T^{2})}{r^{2}B^{2}T^{4}} \Bigg\rbrace^{\frac{1}{2}}t^{\circ}.
\end{eqnarray}

\noindent So far we have not used any theory of gravity as such. One can further relate $R_{\alpha\beta}$ to the energy-momentum tensor $T_{\alpha\beta}$ for a particular theory, thereby arriving at the energy conditions. The two basic issues at hand are: (a) whether the expansion could be always finite and (b) in that case, is the convergence condition violated or satisfied. We will address these points in the 
forthcoming discussion.

We also discuss the necessary and sufficient conditions under which a geometry similar to Eq. (\ref{metricgeodesic1}) can describe the characteristic feature of a wormhole geometry, the throat. The throat is a space-like closed two-dimensional surface of the minimum area. To determine the throat conditions, we construct an embedding geometry for the metric 
\begin{equation}\label{metricgeodesic2}
ds^{2} = T(t)^{2} \Bigg[A(r)^{2}dt^{2} - B(r)^{2}dr^{2} - r^{2}d\Omega^2 \Bigg],
\end{equation}
or
\begin{equation}\label{metricgeodesic3}
ds^{2} = T(t)^{2} \Bigg[A(r)^{2}dt^{2} - \frac{1}{\Big(1 - \frac{b(r)}{r}\Big)A(r)^{2}}dr^{2} - r^{2}d\Omega^2 \Bigg].
\end{equation}

We consider the metric Eq. (\ref{metricgeodesic2}) on the spatial slice of constant $t$ and $\theta=\pi/2$ and write
\begin{equation}\label{3_d}
dl^2 = B^{2}T^{2}dr^{2} + r^{2}T^{2}d\phi^{2}.
\end{equation}
$dl^2$ is the metric on a surface of revolution $\rho = \rho(z)$ embedded in a three-dimensional space with an Euclidean metric
\begin{equation}\label{R3d}
dl^2 = dz^2 + d\rho^2 + \rho^2 d\phi^2,
\end{equation}
where $z$, $\rho$ and $\phi$ are cylindrical coordinates. Comparing Eqs. (\ref{3_d}) and (\ref{R3d}) we get
\begin{eqnarray}
&&\rho^2 = r^{2}T^{2},\\&&
dz^{2} + d\rho^{2} = B^{2}T^{2}dr^{2}.
\end{eqnarray}

For a constant $t$ we can write
\begin{equation}
d\rho = T dr,
\end{equation}
and
\begin{eqnarray}\label{ee2}
&& \frac{d\rho}{dz} = \frac{1}{(B^{2}-1)^{\frac{1}{2}}}, \\&&
\frac{d^2\rho}{dz^2} = -\frac{B'B}{T(B^{2}-1)^{2}}.\label{ee3}
\end{eqnarray}

\noindent The throat of the wormhole has the projected shape of a sphere, located at a certain value of the radial coordinates $r = r_{w}$. On an embedding diagram, this sphere of $r = r_{w}$ simply corresponds to a circle of radius $\rho$ on the surface of revolution. Naturally, at the throat the radius of the circle $\rho(z)$ has a minimum. Conditions for this minimum of $\rho(z)$ at $r=r_{w}$ are
\begin{eqnarray}\label{ee4}
&&\frac{d\rho}{dz}\Big|_{r_{w}} = 0,\\&&
\frac{d^2\rho}{dz^2}\Big|_{r_{w}} > 0.\label{ee5}
\end{eqnarray}
Comparing these two sets of equations, we write the first throat condition for the metric in Eq. (\ref{metricgeodesic2}) as
\begin{eqnarray}\label{ee6}
&&\frac{d\rho}{dz}\Big|_{r_{w}} = \Bigg[\frac{A^{2}(r)(r - b(r))}{r - A^{2}(r)(r - b(r))}\Bigg]^{\frac{1}{2}} \Bigg|_{r_{w}},
\end{eqnarray}
which means that the first throat condition is simply
\begin{equation}\label{ee7}
r_{w} = b(r)|_{r_{w}}.
\end{equation}
Using Eqs. (\ref{ee3}), (\ref{ee5}) and (\ref{ee7}), we write the second throat condition as 
\begin{equation}\label{ee8}
\frac{A^{2}(r)}{2T(t)b(r)} \left\lbrace 1 - b'(r) \right\rbrace \Big|_{r_{w}} > 0.
\end{equation}

\noindent The advantage of the study in this section is that even without an exact solution of the background collapsing geometry, we do realize the allowed dynamics in a general sense and can throw some light on the possible end-state of the collapse. However, having a solution of the field equations makes the narrative even more clear as one can then write $A(r)$, $B(r)$ abd $T(t)$ in some exact form and plot the evolution of the kinematic quantities. In the next section, we first write an exact solution of the collapsing fluid distribution and then study the evolution of expansion and the Null Convergence Condition for the geodesic congruence.

\section{An Exact Collapsing solution and the Simpson-Visser wormhole end-state}
The theory we consider is GR and we write the action as
\begin{equation}\label{action}
A = \int\Bigg(\frac{R}{16\pi G} + L_{m}\Bigg)\sqrt{-g}~d^{4}x.
\end{equation}
Metric variation of this action leads to the Einstein field equations
\begin{equation}\label{fe2}
G_{\mu\nu}= \kappa T^{m}_{\mu\nu}.
\end{equation}
$L_{m}$ defines the fluid contribution and the fluid is assumed to have pressure anisotropy and a radial heat flux. The energy-momentum tensor can therefore be written as
\begin{equation}\label{EMT}
T_{\mu\nu}=(\rho+p_{t})u_{\mu}u_{\nu}-p_{t}g_{\mu\nu}+ (p_r-p_{t})\chi_{\mu}\chi_{\nu} + q_{\mu}u_{\nu} +q_{\nu}u_{\mu}.
\end{equation}
$q^{\mu}=(0,q,0,0)$ is the radial heat flux, $\rho$ the energy density, $p_{t}$ and $p_r$ are the tangential and radial pressure, respectively. $u_{\mu}$ and $\chi_{\mu}$ are the four-velocity and unit space-like radial four-vector. The usual normalizations are given as
\begin{equation}
u^{\mu}u_{\mu}=1,\quad\chi^{\mu}\chi_{\mu}=-1,\quad\chi^{\mu}u_{\mu}=0.
\end{equation}

We begin by writing the metric of the interior of an initially collapsing fluid distribution in the form of Eq. (\ref{metricgeodesic1}) 
\begin{equation}\label{vzero}
ds^2_{-} = T(t)^{2} \Big[A(r)^{2}dt^{2} - B(r)^{2}dr^{2} - r^{2}d\Omega^2 \Big].
\end{equation}

Even before considering the field equations, we deduct the time evolution of the system from the boundary matching condition with a suitable exterior geometry. Since the energy momentum distribution inside the collapsing body has a radial heat flux, we assume the exterior to be given by a Vaidya metric
\begin{equation} \label{v1}
ds^2_{+} =  \left(1 - \frac{2M(u)}{R}\right) du^2 + 2dudR - R^{2}d\Omega^{2}.
\end{equation}
The Newtonian mass of the gravitating body is $M(u)$, in reference with an observer at infinity. The metric in Eq. (\ref{v1}) is the unique spherically symmetric solution of the Einstein field equations for null fluid. We assume that the interior and the exterior geometry are joined through a spherical surface $\Sigma$ (a time-like three-space). We require a smooth matching of these two geometries (for more discussion on matching of fundamental forms, see for instance \cite{israel, santosmatching, chanmatching}) and hence we first demand
\begin{equation}
ds^2_{-} \overset{\Sigma}{=} ds^2_{+},
\label{eq:dsidso}
\end{equation}
which is the matching of metric or the first fundamental form. We also demand the smooth matching of extrinsic curvatures defined at the boundary hypersurface, i.e.,
\begin{equation}
K^{-}_{ij} = K^{+}_{ij},
\label{eq:kijikijo}
\end{equation}
with
\begin{equation}
K^{\pm}_{ij}=-n^{\pm}_{\alpha}
{{\partial^2x^{\alpha}} \over {\partial \xi^i \partial \xi^j}}
-n^{\pm}_{\alpha}\Gamma^{\alpha}_{\beta \gamma}
{{\partial x^{\beta}}  \over {\partial \xi^i}}
{{\partial x^{\gamma}} \over {\partial \xi^j}}.
\label{eq:kij}
\end{equation}
The entities defined on the RHS of extrinsic curvature tensor are
\begin{itemize}
\item {The unit normal vectors to $\Sigma$, $n^{\pm}_{\alpha}$-s,}
\item {Coordinates of two spacetimes, $x^{\alpha}$-s,}
\item {Coordinates on the surface $\Sigma$, $\xi^i$-s,}
\item {The Christoffel symbols, $\Gamma^{\alpha}_{\beta \gamma}$-s.}
\end{itemize}
We also take $\tau$ to be a time coordinate on $\Sigma$ and use Eq. (\ref{eq:dsidso}) to write
\begin{equation}
{\frac{dt}{d\tau}} \overset{\Sigma}{=} \frac{1}{AT},
\label{eq:ts}
\end{equation}
\begin{equation}
rT \overset{\Sigma}{=} R(u),
\label{eq:cs}
\end{equation}
\begin{equation}
\left(\frac{du}{d\tau}\right)^{-2} \overset{\Sigma}{=} \left(1 - \frac{2M}{R} + 2 \frac{dR}{du}. \right)_{\Sigma},
\label{eq:dvdtau}
\end{equation}
$n^{\pm}_{\alpha}$-s are found to be
\begin{equation}
n^{-}_{\alpha} \overset{\Sigma}{=} BT\delta^1_{\alpha},
\label{eq:ni}
\end{equation}
and
\begin{equation}
n^{+}_{\alpha} \overset{\Sigma}{=} \left( 1 - {\frac{2M}{R}} + 2 {\frac{dR}{du}} \right)^{-1/2} \left(-{\frac{dR}{du}}\delta^0_{\alpha} + \delta^1_{\alpha} \right).
\label{eq:no}
\end{equation}
With this, the extrinsic curvature components are derived as
\begin{equation}
K^{-}_{\tau \tau} \overset{\Sigma}{=} -\left[ {\left( {\frac{dt}{d\tau}}\right)}^2 {\frac{A'A}{B}T} \right],
\label{eq:k00i}
\end{equation}
\begin{equation}
K^{-}_{\theta \theta} \overset{\Sigma}{=} \left({\frac{rT}{B}} \right),
\label{eq:k22i}
\end{equation}
\begin{equation}
K^{-}_{\phi \phi} = K^{-}_{\theta \theta} \sin^2 \theta,
\label{eq:k33i}
\end{equation}
\begin{equation}
K^{+}_{\tau \tau} \overset{\Sigma}{=} \left[ {\frac{d^{2}u}{d\tau^2}} {\left(\frac{du}{d\tau}\right)}^{-1} - {\left(\frac{du}{d\tau}\right)} {\frac{M}{R^{2}}} \right],
\label{eq:k00o}
\end{equation}
\begin{equation}
K^{+}_{\theta \theta} \overset{\Sigma}{=} \left[ {\left(\frac{du}{d\tau} \right)} \left(1 - {\frac{2M}{R}} \right)R + {\frac{dR}{d\tau}}{R} \right],
\label{eq:k22o}
\end{equation}
and
\begin{equation}
K^{+}_{\phi \phi} = K^{+}_{\theta \theta} \sin^2 \theta.
\label{eq:k33o}
\end{equation}
We use Eqs. (\ref{eq:k22i}) and (\ref{eq:k22o}) to write
\begin{equation}
\left[ {\left(\frac{du}{d\tau} \right)} \left(1 - {\frac{2M}{R}} \right)R + {\frac{dR}{d\tau}}{R} \right] \overset{\Sigma}{=} \left({\frac{rT}{B}} \right).
\label{eq:k22ik22o}
\end{equation}

We further simplify Eq. (\ref{eq:k22ik22o}) using Eqs. (\ref{eq:ts}), (\ref{eq:cs}) and (\ref{eq:dvdtau}) and write
\begin{equation}
M \overset{\Sigma}{=} \frac{r T}{2}\Bigg[\frac{r^2}{A(r)^2}\frac{\dot{T}^2}{T^2} - \frac{1}{B(r)^2} + 1\Bigg].
\label{eq:ms}
\end{equation}

The function $m(r,t) = \frac{r T}{2}\Big[\frac{r^2}{A(r)^2}\frac{\dot{T}^2}{T^2} - \frac{1}{B(r)^2} + 1\Big]$ is the mass function derived by Misner and Sharp \cite{MS, cahill}. It gives us the energy entrapped within the boundary hypersurface. Now, Eqs. (\ref{eq:k00i}), (\ref{eq:k00o}) and (\ref{eq:ts}) altogether produces
\begin{equation}
\left[ {\frac{d^{2}u}{d\tau^2}} {\left(\frac{du}{d\tau}\right)}^{-1} - {\left(\frac{du}{d\tau}\right)} {\frac{M}{R^2}} \right] \overset{\Sigma}{=} -\left( {\frac{A'}{ABT}} \right).
\label{eq:k00ik00o}
\end{equation}

Moreover, Eqs. (\ref{eq:ts}), (\ref{eq:cs}), (\ref{eq:ms}) and (\ref{eq:k22ik22o}) produce
\begin{equation}
\left( {\frac{du}{d\tau}} \right) \overset{\Sigma}{=} \left(\frac{1}{B} + {\frac{r\dot{T}}{AT}} \right)^{-1},
\label{eq:dvdtau1}
\end{equation}

whose $\tau$ derivative can be used to rewrite Eq. (\ref{eq:k00ik00o}) as
\begin{eqnarray}\nonumber
&& {\Big( \frac{r}{2ABT} \Big)} \Bigg[- 2 \frac{A'\dot{T}}{AT} + \frac{B}{A} \Bigg\lbrace 2{\frac{\ddot{T}}{T}} - \frac{\dot{T}^2}{T^2} \\&&
+ \Bigg(\frac{A^2}{r^2} - \frac{A^{2}}{r^{2}B^{2}} - 2\frac{AA'}{rB^2} \Bigg)\Bigg\rbrace \Bigg] \overset{\Sigma}{=} 0.
\label{eq:pqb}
\end{eqnarray}

Using the above constructions we write the continuity of the metric as
\begin{eqnarray}
&& (rT) \overset{\Sigma}{=} R, \\&&
m(t,r) \overset{\Sigma}{=} M.
\end{eqnarray}

However, this can be also realized by simply writing the interior metric Eq. (\ref{vzero}) in Eddington retarded coordinates,
\begin{eqnarray}\label{vzerozero}\nonumber
&& ds^2_{-} =  \Bigg[1 - \Big(\frac{r^2}{A(r)^2}\frac{\dot{T}^2}{T^2} - \frac{1}{B(r)^2} + 1\Big)\Bigg] du^2 \\&&
+ 2dudR - R^{2}d\Omega^{2}.
\end{eqnarray}

Intriguingly, a generalized metric with similar form has been already considered by Simpson, Martin-Moruno and Visser \cite{simpson2} as a non-static version of the so-called black-bounce scenario. For a collapsing sphere, we take that the $\Sigma$ as a boundary hypersurface located at some radial coordinate $r = r_{b}$. We also define $A_{0} = A(r)_{(r=r_{b})}$, $A_{1} = A'(r)_{(r=r_{b})}$, $B_{0} = B(r)_{(r=r_{b})}$ and $B_{1} = B'(r)_{(r=r_{b})}$. With this, the continuity of extrinsic curvature as in Eq. (\ref{eq:pqb}) becomes a time-dependent differential equation, since the r-dependent functions contribute as constants at the boundary. 
\begin{eqnarray}\nonumber \label{extrinscurv}
&&\frac{2B_{0} \ddot{T}}{A_{0}T} - \frac{B_{0} \dot{T}^2}{A_{0} T^2} - \frac{2A_{1} \dot{T}}{A_{0}T} - \frac{A_{0}}{B_{0}} \Bigg[\frac{2A_{1}}{A_{0}r_{b}} + \frac{1}{r_{b}^2} \\&&
- \frac{B_{0}^2}{r_{b}^2}\Bigg] \overset{\Sigma}{=} 0.
\end{eqnarray}

We write $\frac{2A_{1}}{B_{0}} = \lambda_{1}$ and $\frac{A_{0}^2}{B_{0}^2} \Big[\frac{2A_{1}}{A_{0}r_{b}} + \frac{1}{r_{b}^2} - \frac{B_{0}^2}{r_{b}^2}\Big] = \lambda_{2}$. Therefore, Eq. (\ref{extrinscurv}) becomes
\begin{equation}\label{decase1}
\frac{2\ddot{T}}{T} - \frac{\dot{T}^2}{T^2} - \lambda_{1} \frac{\dot{T}}{T} - \lambda_{2} = 0.
\end{equation}
\begin{figure}[h]
	\begin{center}
		\includegraphics[angle=0, width=0.35\textwidth]{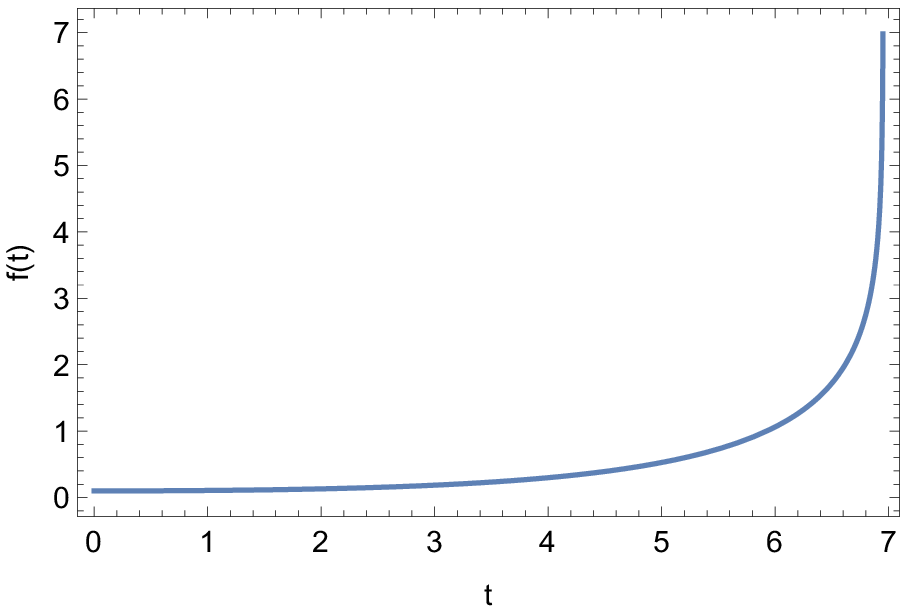}
		\includegraphics[angle=0, width=0.35\textwidth]{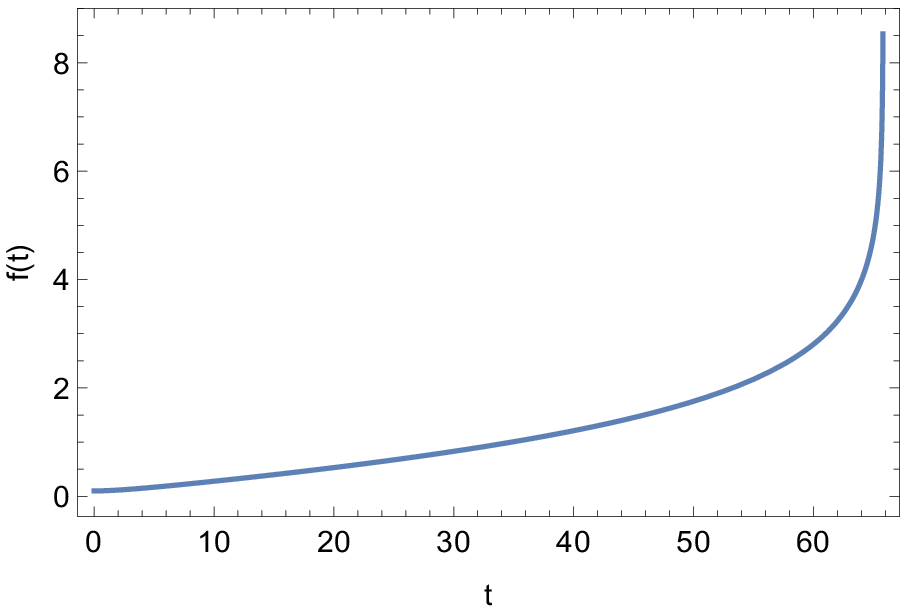}
	\end{center}
	\caption{{\small Evolution of $f(t)$ as a function of $t$. For the graph on top, $\lambda_{1} > 0$ and for the graph below, $\lambda_{1} < 0$. All the other parameters, when varied, scales these behavior while keeing the qualitative feature fixed.}}
	\label{scale_conf0}
\end{figure}

We are interested to see if the above equation has any solution of the form 
\begin{equation}\label{ansatz}
T(t) = m + n e^{\alpha f(t)}.
\end{equation}
The time evolution of the function $f(t)$ will ensure the structure and fate of the collapsing geometry. Using this ansatz in Eq. (\ref{decase1}) we write
\begin{eqnarray}\label{decase2}
&& \ddot{f} = \frac{\left(m + n e^{\alpha f} \lambda_{2}\right)}{2\alpha n e^{\alpha f}} - \\&&\nonumber
\frac{\left\lbrace n \alpha^{2} \dot{f}^{2} e^{(\alpha f)} \left(2m + n e^{\alpha f}\right) - \alpha \lambda_{1} n \dot{f} e^{\alpha f} \left(m + n e^{\alpha f}\right)\right\rbrace}{\left(m + n e^{\alpha f}\right)2\alpha n e^{\alpha f}}.
\end{eqnarray}
We solve Eq. (\ref{decase2}) numerically and study the evolution of $f(t)$. From Eq. (\ref{ansatz}) one can note that for a collapsing solution, $\alpha \frac{df}{dt}$ has to be negative which is one of the constraints over the numerical solution.
\begin{figure}[h]
	\begin{center}
		\includegraphics[angle=0, width=0.35\textwidth]{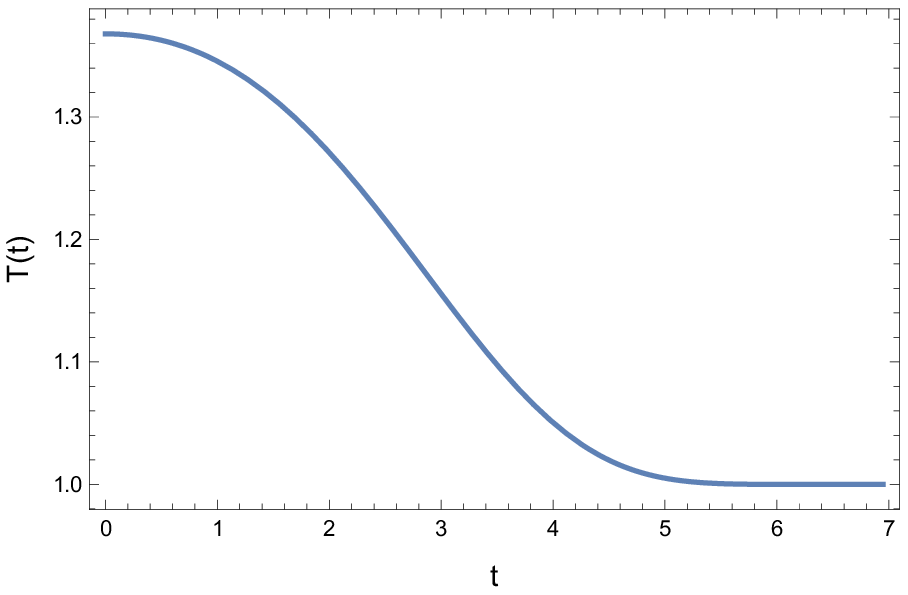}
		\includegraphics[angle=0, width=0.35\textwidth]{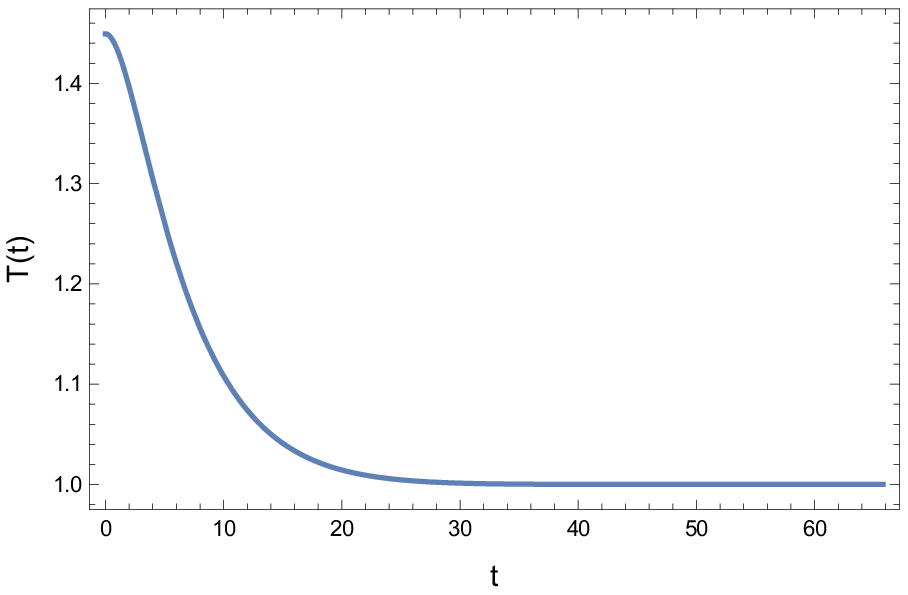}
	\end{center}
	\caption{{\small Evolution of $T(t)$ as a function of $t$. For the graph on top, $\lambda_{1} > 0$ and for the graph below, $\lambda_{1} < 0$. All the other parameters, when varied, scales these behavior while keeping the qualitative feature fixed. We choose $m = n = 1$, and $\alpha = -10$.}}
	\label{scale_conf1}
\end{figure}
Fig. \ref{scale_conf0} shows the time evolution of the function $f(t)$ for $\alpha < 0$. We have chosen $\alpha = -1$ to show a particular case as an example, but a differently chosen negative value of $\alpha$ simply scales the qualitative behavior of the evolution. The same applies for the other parameters as well, $m$ and $n$, which we choose to be positive. $f(t)$ 
has an increasing profile as a function of time and goes to infinity at some finite value of time. The value of $\lambda_1$ determines this time of divergence of $f(t)$. In the plots, we have shown that for $\lambda_1 < 0$, $f(t)$ diverges at a later time than the case where $\lambda_1 > 0$. In our example, we have chosen $\lambda_1 = \pm 1$ and $\lambda_2 = -0.01$ as an example. Therefore, for all $\alpha < 0$, $e^{\alpha f(t)}$ decreases as a function of time. When $f(t) \rightarrow \infty$, $e^{\alpha t} \rightarrow 0$ and therefore $T(t) \sim m$. Therefore, the collapse forms a geometry with a finite non-zero radius of a two-sphere, without reaching a zero proper volume. It is curious that to maintain a collapsing profile, $\lambda_{2}$ must be a negative quantity. For all $\lambda_{2} > 0$, $f(t)$ has a negative evolution, making the scenario non-collapsing. We plot the evolution of the conformal factor $T(t)$ in Fig. \ref{scale_conf1} for $m = n = 1$, and $\alpha = -10$. Thus, the end-result has the form 
\begin{equation}\label{metricconformal00}
ds^{2} = \left(m + n e^{\alpha f(t)}\right)^{2} \Bigg[ A(r)^{2}dt^{2} - B(r)^{2}dr^{2} - r^{2}d\Omega^2 \Bigg].
\end{equation}

Having determined the time evolution independently, we now extrapolate the inhomogeneities in the metric coefficients intuitively. The primary constraint we follow is $\lambda_2 < 0$. We also keep in mind that the radial functions $A(r)$ and $B(r)$ must be so chosen such that the metric has a geometry similar to (\ref{beginning}) as well as at the boundary hypersurface similar to Vaidya. However, a solution written in this manner is not a unique solution as there can be other choices of $A(r)$ and $B(r)$ that satisfy the relevant conditions. This is a special example. We write the solution for the interior of the stellar body i.e., for $r_{w} < r \leq r_{b}$,

\begin{eqnarray}\nonumber \label{metricconformal0}
&&ds^{2} = \left(m + n e^{\alpha f(t)}\right)^{2} \Bigg[\Bigg(1-\frac{x_{1}(r)}{r}\Bigg)dt^{2} \\&&
- \frac{dr^{2}}{\Bigg(1 - \frac{b(r)}{r}\Bigg)\Bigg(1-\frac{x_{2}(r)}{r}\Bigg)} - r^{2} d\Omega^2 \Bigg].
\end{eqnarray}

\begin{eqnarray}\nonumber
&&x_{1}(r) = 2 m_{0} e^{\delta_{1}^{2}(r - r_{b})},\\&& \label{radial1}
b(r) = b_{0}^{2}(r_{b}^{2} - r^{2})/r,\\&& \label{radial2}
x_{2}(r) = 2 m_{1} e^{\delta_{2}^{2}(r - r_{b})}. \label{radial3}
\end{eqnarray}
$r_{w}$ is the allowed minimum value of $r$, as discussed through Eq. (\ref{coordinate}),
\begin{eqnarray}\nonumber
&&r\in(r_{w},+\infty);\\&& 
t\in(0,+\infty).
\end{eqnarray}
At the boundary of the stellar body $r=r_b$,
\begin{eqnarray}\nonumber\nonumber\nonumber
&&x_{1} = 2 m_{0},\\&&
b(r) = 0,\\&&
x_{2} = 2 m_{1}.
\end{eqnarray}

The properties of the collapsing object are summarized below.
\begin{enumerate}
\item{Time evolution of the collapsing object is allowed until $n e^{\alpha f(t)} \rightarrow 0$. No zero proper volume can be reached at any value of time. A static geometry is formed with a wormhole-like structure similar to (\ref{beginning}), with a constant conformal factor $m$.}
\item{The term $\Big(1 - \frac{b(r)}{r}\Big)$ is responsible for the wormhole geometry equivalent to the Simpson-Visser metric as in Eq. (\ref{beginning}). At the boundary $r = r_{b}$, the term $b_{0}^{2}(r_{b}^{2} - r^{2})$ is $0$, which indicates that at the boundary, the wormhole structure is not realized. It comes into picture only inside the collapsing body.}
\end{enumerate}

We note here that the Misner Sharp mass function given in Eq. (\ref{eq:ms}) asymptotically becomes a constant (Schwarzschild) when the collapse reaches its end-state. This is clear from the time evolving conformal factor in Eq. (\ref{ansatz}) which, in the limit $t \rightarrow \infty$ becomes a constant. Moreover, when a Wormhole geometry develops as an end state, $r \rightarrow r_{w}$ which corresponds to the constant value of the radial coordinate at the wormhole throat. Therefore, the matching holds at the end of the collapse as well, with a constant mass function $M = \frac{m r_{w}}{2} \left\lbrace 1 - \left(1 - \frac{b(r_{w})}{r_{w}} \right)\left(1-\frac{x_{2}(r_{w})}{r_{w}}\right)\right\rbrace$. \\

From the Einstein field equations one can write the expression of density, radial and tangential pressure and the radial heat flux. As an example we study these entities for a comoving frame, where the four-velocity field is
\begin{equation}
u^{\mu} = \frac{\delta^{\mu}_{0}}{\sqrt{g_{00}}}.
\end{equation}

\begin{figure}[t]
	\begin{center}
		\includegraphics[angle=0, width=0.35\textwidth]{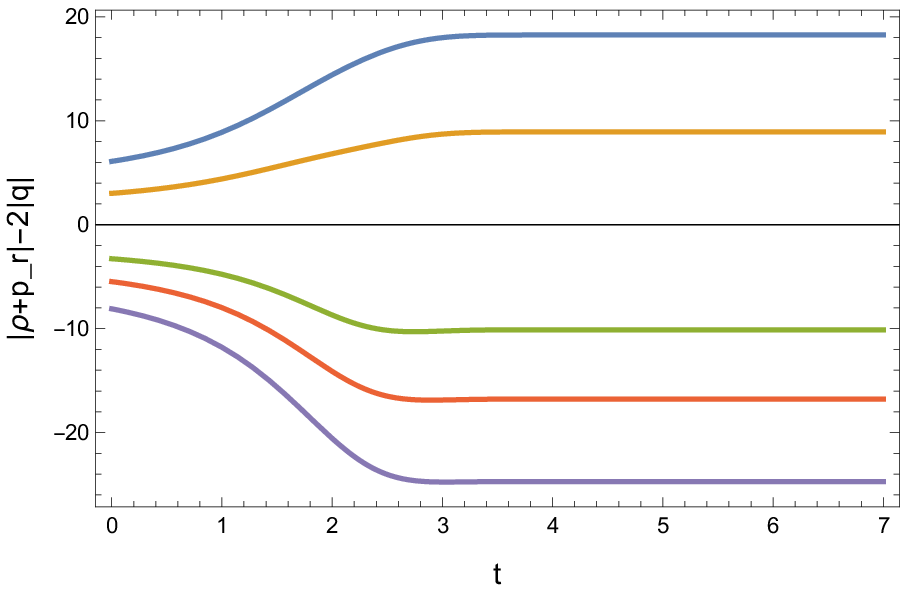}
		\includegraphics[angle=0, width=0.35\textwidth]{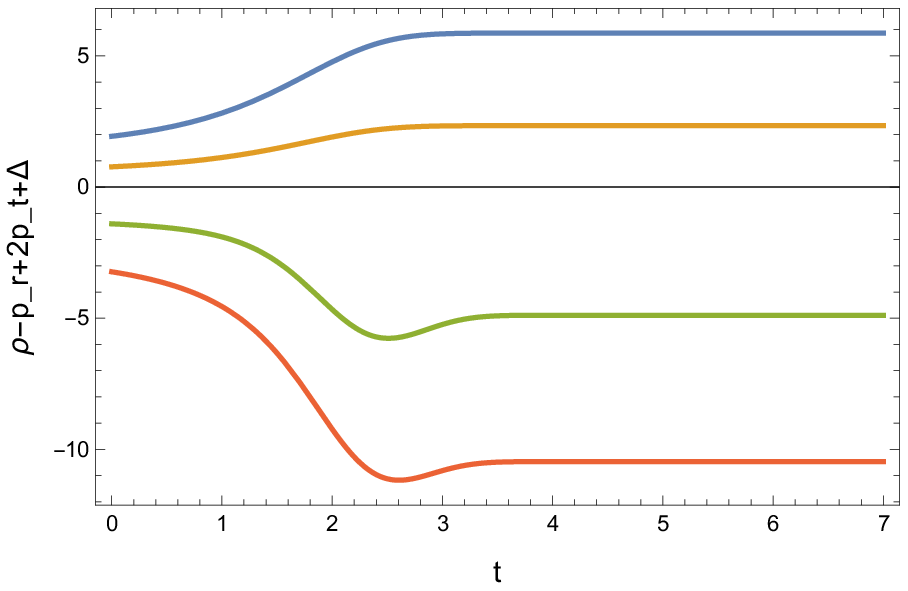}
		\includegraphics[angle=0, width=0.35\textwidth]{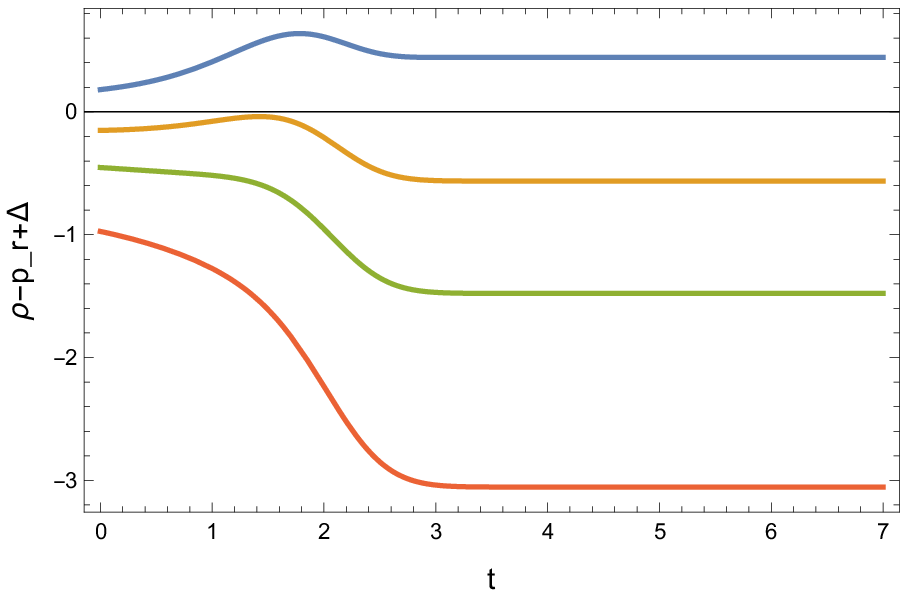}
		\includegraphics[angle=0, width=0.35\textwidth]{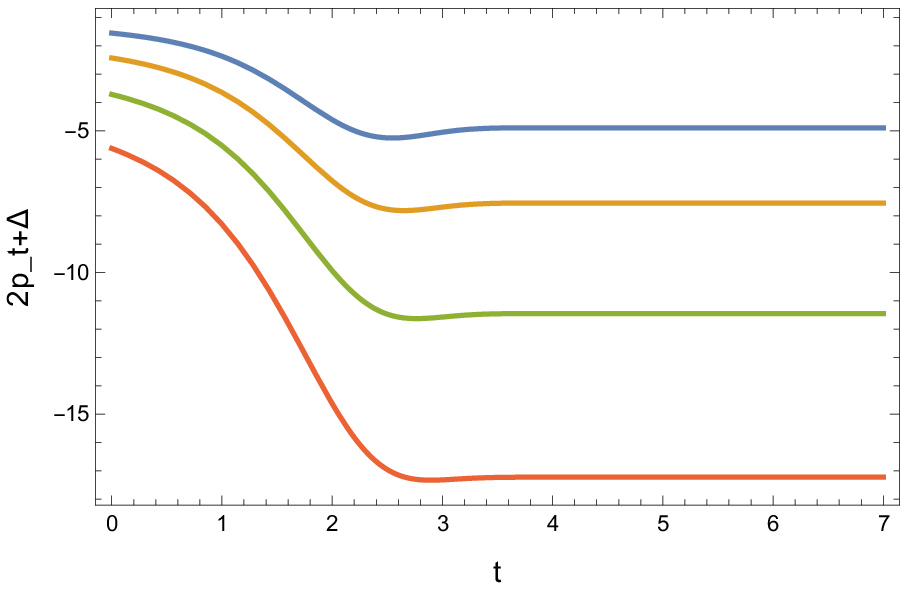}
	\end{center}
	\caption{{\small Energy conditions as a function of time for the solutions in Eq. (\ref{metricconformal0}), (\ref{radial1}), (\ref{radial2}) and (\ref{radial3}). From top to bottom in sequence the graphs show the plots of $NEC1$, $NEC2$, $WEC$ and $SEC$. Different colored curves in a single graph denote the evolution for different radial distance from the centre; $r \sim 0.98 (Blue)$, $r \sim 0.9 (Yellow)$, $r \sim 0.8 (Green)$ and $r \sim 0.7 (Green)$. The other parameters are chosen as $m = n = 1$, $\alpha = -10$, $m_{0} = 0.1$, $m_{1} = 0.11$, $\delta_{1} = 0.2$, $\delta_{2} = 0.202$, $b_{0} = 2$ and $r_{b} = 1$.}}
	\label{EC_conf1_0}
\end{figure}

\begin{equation}
\rho = \frac{3\dot{T}^{2}}{A^{2}T^{4}} - \frac{1}{B^{2} T^{2}}\Bigg[\frac{1}{r^2} - \frac{2B'}{rB} - \frac{B^{2}}{r^2}\Bigg].
\end{equation}

\begin{equation}
p_{r} = -\frac{1}{A^{2} T^{2}} \Bigg[\frac{2\ddot{T}}{T} - \frac{\dot{T}^2}{T^2}\Bigg] + \frac{1}{B^{2} T^{2}}\Bigg[\frac{2A'}{Ar} + \frac{1}{r^2}\Bigg] - \frac{1}{r^{2} T^{2}}.
\end{equation}

\begin{equation}
p_{t} = -\frac{1}{A^{2} T^{2}} \Bigg[\frac{2\ddot{T}}{T} - \frac{\dot{T}^2}{T^{2}}\Bigg] + \frac{1}{B^{2}T^{2}} \Bigg[\frac{A''}{A} - \frac{A'B'}{AB} + \frac{A'}{rA} - \frac{B'}{rB}\Bigg].
\end{equation}

\begin{equation}\label{heatfluxconf}
q = -\frac{2\dot{T}A'}{A^{2}B^{2}T^{4}}.
\end{equation}

Satisfying the energy conditions in the gravitational collapse of a fluid distribution essentially imply  the requirement of a `locally positive energy density' everywhere. We are aware of different versions 
such as the Null Energy Condition, the Weak Energy Condition and the Strong Energy Condition. Algebraically they are written as follows: 

\begin{eqnarray}\label{ec1}
&NEC1 \rightarrow \mid \rho + p_{r} \mid- 2\, \mid q\mid \geq 0 ~,\\
&NEC2 \rightarrow \rho - p_{r} + 2\,p_{t}+\bigtriangleup \geq 0 ~,\\
&WEC \rightarrow \rho - p_{r} +\bigtriangleup \geq 0 ~,\\
&SEC \rightarrow 2\,p_{t}+ \bigtriangleup \geq 0 ~,\\
&\bigtriangleup = \sqrt{(\rho + p_{r})^2 - 4\,q^2}.
\end{eqnarray}

For a detailed discussion of the energy conditions for an imperfect fluid we refer to the work of Kolassis, Santos and Tsoubelis \cite{kolassis}, Pimentel, Lora-Clavijo and Gonzalez \cite{pimentel}. To plot the energy conditions as a function of time we choose $m = n = 1$, $\alpha = -10$, $m_{0} = 0.1$, $m_{1} = 0.11$, $\delta_{1} = 0.2$, $\delta_{2} = 0.202$, $b_{0} = 2$ and $r_{b} = 1$ and plot the LHS of the conditions in Eq. (\ref{ec1}). Ideally, $m_{0} = m_{1}$ and $\delta_{1} = \delta_{2}$ define a special case and we take them to be slightly different from each other for some additional generality. Fig. \ref{EC_conf1_0} suggests that the energy conditions near the boundary of the sphere $r \sim r_{b}$ can be satisfied. However, they are clearly violated inside the collapsing sphere for $r < r_{b}$. In simple terms, the fluid distribution abides by the energy conditions as long as the wormhole geometry does not form (At $r = r_{b}$ the wormhole function is zero). As soon as $r$ differs reasonably from $r_{b}$ the energy conditions are violated. This coincides with the collapse facing some geometric resistance. Different curves in a single graph show the evolution for different values of the radial coordinate $r$, for instance, the blue curve in all the graphs indicate evolution of the energy conditions for a shell very close to the boundary hypersurface ($r \sim 0.98$), followed by subsequent collapsing shells. yellow for $r \sim 0.95$, green for $r \sim 0.8$ and red for $r \sim 0.7$. \\

However, it is also possible to comment on the scenario in a more general way, using the exact solution in Eq. (\ref{metricconformal0}), the geodesic Eqs. (\ref{eq:geodesic_r}) and (\ref{eq:geodesic_t}) and the expression for $\Theta$ as in Eq. (\ref{thetageneral1}). The evolution of $\Theta$ can be studied for the specific example by using the exact evolution of metric components and writing $\frac{dt}{d\tau}$ as a function of time. This demonstrates whether the expansion can reach $-\infty$ or not, the violation/satisfaction of the Null Convergence Condition (NCC) and associated consequences for the energy conditions and their correlation with a non-singular end-state. \\

\begin{figure}[h]
	\begin{center}
		\includegraphics[angle=0, width=0.35\textwidth]{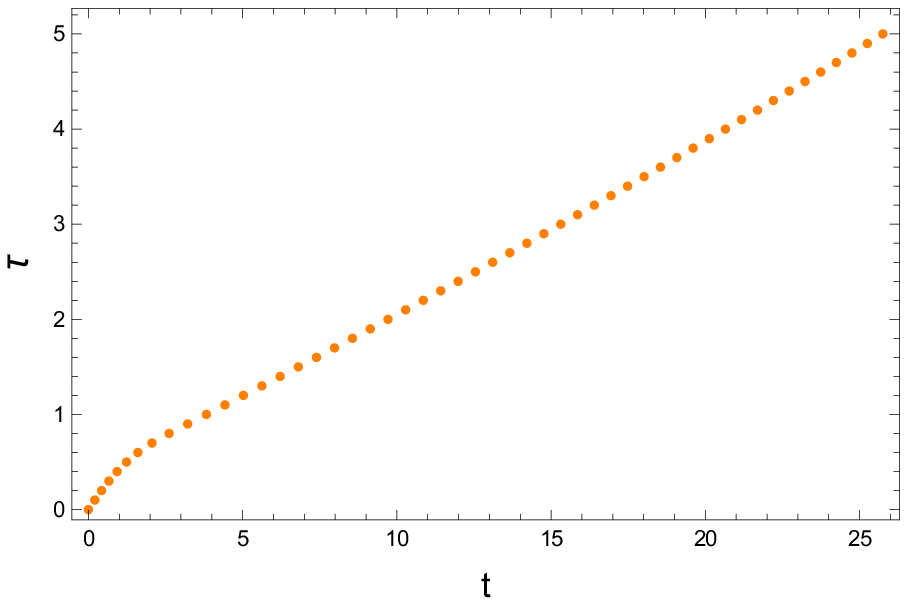}
		\includegraphics[angle=0, width=0.35\textwidth]{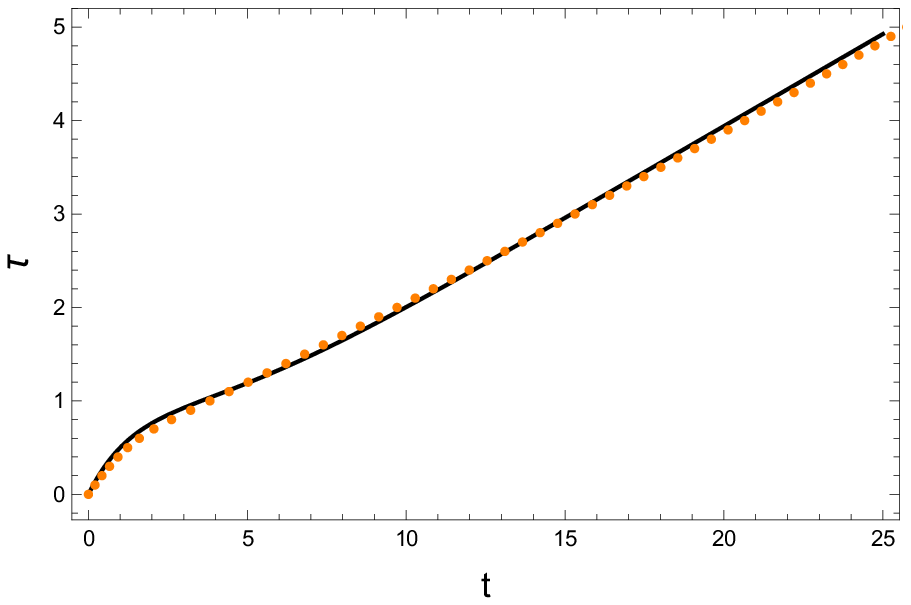}
	\end{center}
	\caption{{\small Evolution of $\tau$ as a function of $t$. $m = n = 1$, $m_{0} = 0.1$, $m_{1} = 0.11$, $\delta_{1} = 0.2$, $\delta_{2} = 0.202$, $b_{0} = 2$ and $r_{b} = 1$. The fitting is done for $r \sim 0.98$, $\tau = \tau_{0}t + \tau_{1}te^{-\tau_{2}t}$, where $\tau_{0} = 0.2$, $\tau_{1} = \tau_{2} = 0.5$.}}
	\label{fit}
\end{figure}

For the geodesic congruence, $u^\alpha=(t^{\circ},r^{\circ},\phi^{\circ})$ where '$\circ$' is the derivative with respect to affine parameter $\tau$. In order to write $t(\tau)$ as a function of $\tau$ we solve the geodesic Eq. (\ref{eq:geodesic_t}) numerically using the exact solution for the metric. The initial condition used to solve the equation is $t(0) = 0$ and $\frac{dt}{d\tau} \sim O(1)$. For a differently chosen initial condition the solution can be different, but we believe this will not affect the qualitative nature of the congruence. In the numerical solution, one needs to use a value of $r$, i.e., one solves for a particular shell of a collapsing sphere labelled by $r$. Of course, different values of $r$ lead to slightly different solutions, however, we demonstrate here for only one case, for $r \sim 0.98$, i.e., a shell very close to the boundary hypersurface. For smaller values of $r$ the $t$ vs $\tau$ nature is just scaled, keeping the qualitative nature unchanged. We plot $\tau$ as a function of $t$, first as data-points and then fit the data using an analytical form in Fig. \ref{fit}.

\begin{figure}[h]
	\begin{center}
		\includegraphics[angle=0, width=0.35\textwidth]{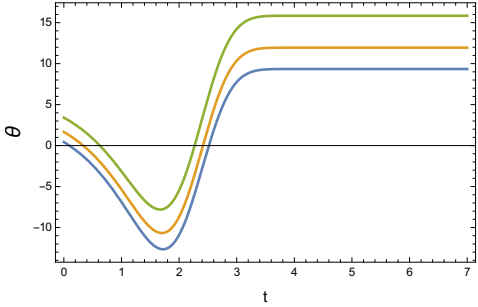}
	\end{center}
	\caption{{\small Evolution of $\Theta$ as a function of $t$ for geodesic congruence. $m = n = 1$, $m_{0} = 0.1$, $m_{1} = 0.11$, $\delta_{1} = 0.2$, $\delta_{2} = 0.202$, $b_{0} = 2$ and $r_{b} = 1$. Curves of different color show the evolution for different values of $r$ (Blue curve $\rightarrow r = 0.95$, Yellow curve $\rightarrow r = 0.85$, Green curve $\rightarrow r = 0.75$ and Red curve $\rightarrow r = 0.65$).}}
	\label{geocongruence1}
\end{figure}

\begin{figure}[h]
	\begin{center}
		\includegraphics[angle=0, width=0.35\textwidth]{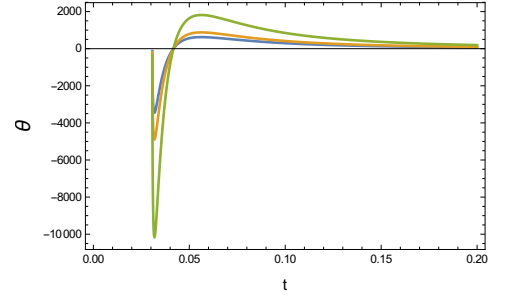}
	\end{center}
	\caption{{\small Evolution of $\Theta$ as a function of $r$ for geodesic congruence. $m = n = 1$, $m_{0} = 0.1$, $m_{1} = 0.11$, $\delta_{1} = 0.2$, $\delta_{2} = 0.202$, $b_{0} = 2$ and $r_{b} = 1$. Curves of different color show the evolution for different values of $t$ (Blue curve $\rightarrow t = 0.1$, Yellow curve $\rightarrow t = 0.5$ and Green curve $\rightarrow t = 1$.}}
	\label{geocongruence2}
\end{figure}

We interpret from the graph that the affine parameter $\tau$ is an increasing function of time $t$ and with time their relationship becomes linear. This can be also explained by using the fitted function which is of the form 
\begin{equation}
\tau = \tau_{0}t + \tau_{1}te^{-\tau_{2}t},
\end{equation}
where $\tau_{0}$, $\tau_{1}$ and $\tau_{2}$ are positive constants. For $r \sim 0.98$ the parameter values of $\tau_{0} = 0.2$, $\tau_{1} = \tau_{2} = 0.5$ give the best fit. For other values of $r$, i.e., for different collapsing shells, similar fitting can be obtained through a similar exercise. As time increases, the exponential factor dies down and one can use $t^{\circ} \sim constant$ for large $t$. With this, we plot the expansion from Eq. (\ref{thetageneral1}) in Fig. \ref{geocongruence1}, using the exact functional form of $T(t)$, $A(r)$ and $B(r)$. We choose the relevant parameters to be $m = n = 1$, $\alpha = -10$, $m_{0} = 0.1$, $m_{1} = 0.11$, $\delta_{1} = 0.2$, $\delta_{2} = 0.202$ and $b_{0} = 2$. Different colors as usual show the evolution for different shells of the sphere labelled by different values of $r$. There is no zero proper volume singularity and $\Theta$ never reaches $-\infty$. Eventually the variation of $\Theta$ for different collapsing shell-values of $r$ slows down, indicating the formation of a stable spherical wormhole-like geometry as an end-state. One can confirm that for very low value of the radial coordinate $r$ ($r << r_{b}$), $\Theta$ is imaginary and evolution is disallowed.  
We also plot the distribution of $\Theta$ as a function of $r$, for different values of time $t$ in Fig. \ref{geocongruence2}. There is no $\Theta$ around $r \sim 0$. More interestingly, the evolution has a remarkable similarity with the static $\Theta$ plot as in Fig. \ref{theta_static}. The departures in the static Simpson-Visser metric in Eq. (\ref{beginning}) and the constructed form Eq. (\ref{metricconformal0}) is in the functional form of $m(r)$. They coincide around the boundary hypersurface $r = r_{b} = 1$, as confirmed by the graphs, however, evolution for small $r$ shows interesting departures. These suggest that the spherical body around the region of wormhole throat formation, goes into a collapsed phase with time, and around the boundary coincides with a standard static Simpson-Visser geometry.

\begin{figure}[h]
	\begin{center}
		\includegraphics[angle=0, width=0.35\textwidth]{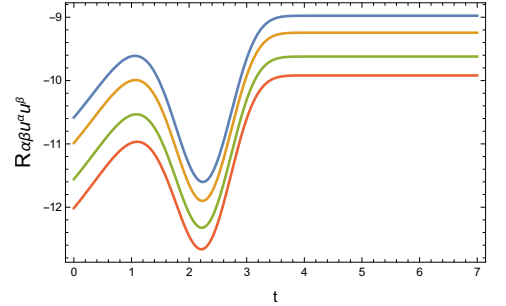}
	\end{center}
	\caption{{\small Evolution of $R_{\alpha\beta}u^\alpha u^\beta$ as a function of $t$ for a geodesic congruence.}} 
	\label{nccfig}
\end{figure}

Using the LHS of Eq. (\ref{nccdetail}), we plot the convergence condition to check if $R_{\alpha\beta}u^\alpha u^\beta \geq 0$ is satisfied or violated. We plot $R_{\alpha\beta}u^\alpha u^\beta$ as a function of time for a geodesic congruence in Fig. \ref{nccfig} and note that for a geodesic congruence, the convergence condition is always violated.

To summarize the results, a collapsing object having a geometry similar to the metrics in Eqs. (\ref{metricconformal00}) or Eq. (\ref{metricconformal0}), containing an imperfect fluid distribution and violating the Null Convergence Condition will collapse and form a wormhole geometry as in Eq. (\ref{simpsonvisser}) or Eq. (\ref{beginning}). There will be no zero proper volume at any time. It is straightforward to check that the Ricci and Kretschmann curvature scalars are finite at all times.  \\

A brief discussion regarding the exterior of the spherical collapsing body is also in the order. As discussed, the exterior is a Vaidya geometry of an outgoing radial flux. We discuss a few things assuming that the exterior is a generalized Vaidya geometry 
\begin{equation}
ds^{2} = \left[1 - \frac{2M(u, R)}{R}\right]du^{2} - 2 \epsilon dudR - r^{2} d\Omega^{2},\;\;(\epsilon = \pm 1).
\end{equation}
$M(u, R)$ is the mass function, the energy entrapped within $R$. $\epsilon = \pm 1$. For an Eddington advanced time $u$, $\epsilon = 1$ and for Eddington retarded time $\epsilon = - 1$. We write the Einstein tensor components for this line element as
\begin{eqnarray}
&& G^{0}_{0} = G^{1}_{1} = - \frac{2 M'(u, R)}{R^{2}}, \\&&
G^{1}_{0} = \frac{2 \dot{M}(u, R)}{R^{2}}, \\&&
G^{2}_{2} = G^{3}_{3} = - \frac{ M''(u, R)}{R},
\end{eqnarray}
where $\dot{M}(u, R) \equiv \frac{\partial M(u, R)}{\partial u}$ and $M'(u, R) \equiv \frac{\partial M(u, R)}{\partial R}$. With this, one is able to write the energy momentum distribution (for a more detailed discussion see \cite{vaidyanewref}) as

\begin{equation}\label{newvaidyaT}
T_{\mu\nu} = T^{(n)}_{\mu\nu} + T^{(m)}_{\mu\nu}.
\end{equation}

The two different components of Energy Momentum tensor are defined as
\begin{eqnarray}
&& T^{(n)}_{\mu\nu} = \mu l_{\mu}l_{\nu},\\&&
T^{(m)}_{\mu\nu} = (\rho + P) \left(l_{\mu}n_{\nu} + l_{\nu}n_{\mu}\right) + P g_{\mu\nu}.
\end{eqnarray}
The tensor components are
\begin{eqnarray}
&& \mu = \frac{2 \epsilon \dot{M}(u, R)}{\kappa R^{2}}, \\&&
\rho = \frac{ 2 M'(u, R)}{\kappa R^{2}}, \\&&
P = - \frac{ M''(u, R)}{\kappa R}.
\end{eqnarray}
$l_{\mu}$ and $n_{\mu}$ are two null vectors with the following definition
\begin{eqnarray}
&& l_{\mu} = \delta^{0}_{\mu}, \\&& 
n_{\mu} = \frac{1}{2}\left[1 - \frac{2M(u, R)}{R}\right]\delta^{0}_{\mu} - \epsilon \delta^{1}_{\mu}, \\&&
l_{\lambda}l^{\lambda} = n_{\lambda}n^{\lambda} = 0, \\&&
l_{\lambda}n^{\lambda} = - 1.
\end{eqnarray}
It is quite interesting to note that $T^{(n)}_{\mu\nu}$ effectively denotes a matter field moving along the  $u = Constant$ null hypersurface. For a general case one usually writes the energy momentum tensor from this, using an orthonormal basis \cite{wangvaidya} consisting of the four vectors  
\begin{eqnarray}
&& E_{0}^{\mu} = \frac{l_{\mu} + n_{\mu}}{\sqrt{2}}, \\&&
E_{1}^{\mu} = \frac{l_{\mu} - n_{\mu}}{\sqrt{2}}, \\&&
E_{2}^{\mu} = \frac{1}{r}\delta^{\mu}_{2}, \\&&
E_{3}^{\mu} = \frac{1}{r\sin\theta}\delta^{\mu}_{3}.
\end{eqnarray}
This leads to the standard energy momentum tensor representation of the exterior as
\begin{equation}
T_{ab} = \left[
\begin{array}{lccl}
\frac{\mu}{2} + \rho& \frac{\mu}{2}& 0 & 0\\
\frac{\mu}{2} & \frac{\mu}{2} - \rho & 0 & 0\\
0 & 0 & P & 0\\
0 & 0 & 0& P\\
\end{array} \right],
\end{equation}
An energy momentum distribution of this type usually signify Type II fluids \cite{hawkingellis}. Energy conditions for these fluids have a standard form which are satisfied or violated depending on the mass function of the system under consideration. The Weak and Strong Energy Conditions are
\begin{equation}
\mu \geq 0, \;\;\; \rho \geq 0, \;\;\; P \geq 0,\; (\mu \neq 0),
\end{equation}
whereas The Dominant Energy Conditions are
\begin{equation}
\mu \geq 0, \;\;\; \rho \geq P \geq 0,\; (\mu \neq 0).
\end{equation}

The present manuscript is entirely devoted to the special case $M(u, R) = M(u)$ and therefore the energy conditions all simplify to produce the necessary condition
\begin{equation}
\mu \sim -\frac{2\frac{dM(u)}{du}}{\kappa R^{2}} \geq 0.
\end{equation}

This is the condition that must be satisfied in the exterior of the collapsing fluid, in order to satisfy the Energy conditions outside the sphere. \\

We conclude this section with a note on the special case $A = 1$, for which the nature of geodesic congruence and the expansion were briefly discussed in Sec. III. As the exact collapsing time evolution is solely based on a smooth boundary matching with the exterior, one needs to write the proper matching condition for this special case as well. Putting $A_{0} = A(r)_{(r=r_{b})} = 1$ and $A_{1} = A'(r)_{(r=r_{b})} = 0$ in Eq. (\ref{extrinscurv}), the matching of extrinsic curvature for this special case produces a differential equation of the form
\begin{equation}
\frac{2\ddot{T}}{T} - \frac{\dot{T}^2}{T^2} + \frac{1}{r_{b}^2}\left(1 - \frac{1}{B_{0}^2} \right) = 0,
\end{equation}
which leads to either an oscillatory or a bouncing solution for $T(t)$ (instead of collapsing) depending on the sign of $\frac{1}{r_{b}^2}\left(1 - \frac{1}{B_{0}^2} \right) = g$. For all $g > 0$, the solution is oscillatory and is not of interest to us. However, for all $g < 0$, the sphere goes through a bounce after an initial phase of collapse, with $T(t) \sim cosh \left[\frac{g^{1/2}}{2}(t + k)\right]^{2}$. This too is an example of a non-singular collapse, however, with a different end-state, as during the bounce the collapsing fluid description disperses away and no static wormhole geometry is reached. 
  
\section{Conclusion}
There are unresolved questions regarding the final fate of a continually collapsing massive star which is among the overall motivations of the present article. We propose a non-singular model of gravitational collapse exploring a different possibility, that of the formation of a wormhole. Although wormhole solutions enjoy their fair share of popularity in gravitational physics, a time-evolving wormhole solution
(though discussed earlier in \cite{sk}) has not been considered in the context of gravitational collapse. Our work shows that even a time-dependent wormhole geometry can 
arise in gravitational collapse, provided the collapsing fluid violates the relevant energy conditions. The wormhole structure is similar to the recently proposed vacuum solution of Simpson and Visser \cite{simpsonvisser} as defined by the metric in Eq. (\ref{simpsonvisser}) or a coordinate transformed version Eq. (\ref{beginning}) of the same. We assume that the collapsing object has a non-trivial time evolution conformal to the static Simpson-Visser geometry and prove that it can collapse forever without reaching any zero proper volume singularity. \\

We study the evolution using kinematic variables, especially the expansion, defined as the rate of change of the cross-sectional area orthogonal to a family of geodesics. We write the evolution equation of these quantities along a geodesic congruence and use the Raychaudhuri equation. The nature of the expansion of the congruence suggests the fate of the collapse. Even without an explicit solution of the field equations, the Raychaudhuri equation helps us to write the Null Convergence Condition, and check it's violation (as well as the violation of the energy conditions) during the collapse. We also give an explicit solution of the field equations with a comoving four-velocity vector. We discuss that the term responsible for a wormhole structure comes from the $g_{11}$ component of the metric and the wormhole geometry is not realized at the boundary hypersurface $r = r_{b}$ of the collapsing sphere. This also inspires the fact that most of the energy conditions are violated inside the collapsing sphere, but not around the boundary hypersurface. We also stress upon the fact that due to this non-trivial structure, the collapsing sphere generates a non-trivial four-acceleration which contributes in resisting the collapse and a formation of zero proper volume.  \\

The solutions are written simply from the boundary matching condition of the two metrics at a suitable boundary surface. This enables one to avoid any assumption of equation of state at the outset. The field equations can be used to study the radiation, density or pressure components in a completely general manner. Moreover, the structure of this solution is quite general as long as the exterior of a collapsing metric is given by a radiating Vaidya solution. Albeit, for a different theory one would require additional constraints on the metric components to be satisfied for consistency. For example, in an $f(R)$ theory a similar solution may be valid along with two additional matching conditions on the Ricci scalar and its normal derivative. Such a generalization can indeed be interesting and will be addressed elsewhere.


\begin{thebibliography}{99}
\bibitem{chandra} S. Chandrasekhar, Astrophys. J. {\bf 74}, 82 (1931).

\bibitem{landau} L. D. Landau, Phys. Z. Sowjetunion {\bf 1}, 285 (1932).

\bibitem{eddi} A. S. Eddington, Mon. Not. Roy. Astron. Soc. {\bf 95}, 194 (1935).

\bibitem{datt} B. Datt, Z. Physik. {\bf 108}, 314 (1938).

\bibitem{oppenvol} J. R. Oppenheimer and G. M. Volkoff, Phys. Rev. {\bf 55}, 374 (1939).

\bibitem{oppensny} J. R. Oppenheimer and H. S. Snyder, Phys. Rev. {\bf 56}, 455 (1939).

\bibitem{penrose} R. Penrose, Nuovo Cimento Rivista Serie 1 (1969).

\bibitem{yod} P. Yodzis, H. J. Seifert, and H. Muller zum Hagen, Comm. Math. Phys. {\bf 34}, 135 (1973) ; Comm. Math. Phys. {\bf 37}, 29 ((1974)).

\bibitem{eard} D. M. Eardley and L. Smarr, Phys. Rev. D. {\bf 19}, 2239 (1979).

\bibitem{christ} D. Christodoulou, Comm. Math. Phys. {\bf 93}, 171 (1984).

\bibitem{joshi} P. S. Joshi, Global Aspects in Gravitation and Cosmology. (Oxford: Clarendon Press,
Oxford University Press. (1993). P. S. Joshi and I. H. Dwivedi, Comm. Math. Phys. {\bf 146}, 333 (1992) ; Phys. Rev. D. {\bf 47}, 5357 (1993) ; Lett. Math. Phys. {\bf 27}, 235 (1993).

\bibitem{waugh} B. Waugh and K. Lake, Phys. Rev. D. {\bf 40}, 2137 (1989) ; Phys. Rev. D. {\bf 38}, 1315 (1988).

\bibitem{ori} A. Ori and T. Piran, Phys. Rev. Lett. {\bf 59}, 2137 (1987) ; Phys. Rev. D. {\bf 42}, 1068 (1990).

\bibitem{joshietal} P. S. Joshi, N. Dadhich, and R. Maartens, Mod. Phys. Lett. A. {\bf 15}, 991 (2000) ; Phys. Rev. D. {\bf 65}, 101501 ((2002)). P. S. Joshi, R. Goswami, and N. Dadhich, Phys. Rev. D. {\bf 70}, 087502 (2004).

\bibitem{chop} M. W. Choptuik, Phys. Rev. Lett. {\bf 70}, 9 (1993).

\bibitem{critical} P. R. Brady, C. M. Chambers, and S. M. C. V. Goncalves, Phys. Rev. D. {\bf 56}, R6057
(1997) ; C. Gundlach, Liv. Rev. Rel. {\bf 2}, 1 (1999) ; C. Gundlach, Phys. Rev. Lett. {\bf 75}, 3214 (1995) ; M. D. Roberts, Gen. Rel. Grav. {\bf 21}, 907 (1989) ; A. V. Frolov, Phys. Rev. D. {\bf 56}, 6433 (1997).

\bibitem{ltb} G. Lemaitre, Ann. Soc. Sci. Bruxelles p. 51 (1933) ; R. C. Tolman, Proc. Natl. Acad. Sci. USA. {\bf 20}, 169 (1934) ; H. Bondi, Mon. Not. R. Astron. Soc. {\bf 107}, 410 (1948).

\bibitem{ritu} R. Goswami, A. M. Nzioki, S. D. Maharaj, and S. G. Ghosh, Phys. Rev. D. {\bf 90}, 084011
(2014).

\bibitem{scfr} S. Chakrabarti, R. Goswami, S. Maharaj and N. Banerjee, Gen. Relativ. Gravit. {\bf 50} : 148 (2018).

\bibitem{scsegb} S. Chakrabarti, Eur. Phys. J. C. {\bf 78} : 296 (2018).

\bibitem{morris1} M. S. Morris and K. S. Thorne, Am. J. Phys. {\bf 56}, 395 (1988).

\bibitem{visser1} M. Visser, Lorentzian Wormholes: From Einstein to Hawking, AIP, College Park, MD, (1995).

\bibitem{flamm} L. Flamm, Phys. Z. {\bf 17}, 448 (1916).

\bibitem{einsteinrosen} A. Einstein and N. Rosen, Ann. Phys. {\bf 2} 242 (1935).

\bibitem{misnerwheeler} C. W. Misner and J. A. Wheeler, Ann. Phys. {\bf 2}, 525 (1957).

\bibitem{hawking} S. W. Hawking and G. F. R. Ellis, The Large Scale Structure of Spacetime Cambridge Uni-
versity Press, Cambridge, England, (1975).

\bibitem{kolassis} C. A. Kolassis, N. O. Santos and D. Tsoubelis, Class. Quant. Gravit. {\bf 5}, 1329 (1988).

\bibitem{pimentel} O. M. Pimentel, F. D. Lora-Clavijo, G. A. Gonzalez : Gen. Relativ. Gravit. {\bf 48}, 124 (2016).

\bibitem{ellis} H. Ellis, J. Math. Phys. {\bf 14}, 104 (1973).

\bibitem{bronnikov} K. A. Bronnikov, Acta Phys. Pol. B. {\bf 4}, 251 (1973).

\bibitem{capo} S. Capozziello, F. S. N. Lobo, and J. P. Mimoso, Phys. Rev. D. {\bf 91}, 124019 (2015) ; Phys. Lett. B. {\bf 730}, 280 (2014).

\bibitem{sk1} B. Bhawal and S. Kar, Phys. Rev. D. {\bf 46}, 2464 (1992).

\bibitem{maeda} H. Maeda and M. Nozawa, Phys. Rev. D. {\bf 78}, 024005 (2008).

\bibitem{sk2} S. Kar, S. Lahiri, and S. SenGupta, Phys. Lett. B. {\bf 750}, 319 (2015).

\bibitem{morris2} M. S. Morris, K. S. Thorne, and U. Yurtsever, Phys. Rev. Lett. {\bf 61}, 1446 (1988).

\bibitem{novikov} I. D. Novikov, Sov. Phys. JETP {\bf 68}, 439 (1989).

\bibitem{frolov} V. P. Frolov and I. D. Novikov, Phys. Rev. D. {\bf 42}, 1057 (1990).

\bibitem{literature} S. V. Sushkov and S. W. Kim, Gen. Relativ. Gravit. {\bf 36}, 1671 (2004) ; S. V. Sushkov, Phys. Rev. D. {\bf 71}, 043520 (2005) ; F. Lobo, Phys. Rev. D. {\bf 71}, 084011 (2005); {\bf 71}, 124022 (2005) ; F. S. N. Lobo and M. A. Oliveira, Phys. Rev. D. {\bf 80}, 104012 (2009) ; M. H. Dehghani and S. H. Hendi, Gen. Relativ. Gravit. {\bf 41}, 1853 (2009) ; P. Kanti, B. Kleihaus, and J. Kunz, Phys. Rev. Lett. {\bf 107}, 271101 (2011) ; T. Harko, F. S. N. Lobo, M. K. Mak, and S. V. Sushkov, Phys. Rev. D. {\bf 87}, 067504 (2013) ; R. Shaikh and S. Kar, Phys. Rev. D. {\bf 94}, 024011 (2016).

\bibitem{simpsonvisser} A. Simpson and M. Visser, JCAP, {\bf 1902} 042 (2019).

\bibitem{roman} T. A. Roman and P. G. Bergmann, Phys. Rev. D. {\bf 28}, 1265 (1983).

\bibitem{Bardeen:1968} J. M. Bardeen, Proceedings of International Conference GR5, Tbilisi, USSR, p. 174., (1968).

\bibitem{Bergmann-Roman} T. A. Roman and P. G. Bergmann, Phys. Rev. D. {\bf 28}, 1265 (1983).

\bibitem{Hayward:2005} S. A. Hayward, Phys. Rev. Lett. {\bf 96}, 031103 (2006).

\bibitem{rc} A. Raychaudhuri. Phys. Rev., {\bf 98}(4) : 1123, (1955).

\bibitem{congrusk} A. Dasgupta, H. Nandan and S. Kar, Phys. Rev. D. {\bf 79}, 124004 (2009) ; S. Ghosh, A. Dasgupta and S. Kar, Phys. Rev. D. {\bf 83}, 084001 (2011).

\bibitem{rajibulsk} R. Shaikh, S. Kar and A. DasGupta, Phys. Rev. D. {\bf 90}, 124069 (2014).

\bibitem{metricchoice} M. M. Som, and N. O. Santos, Phys. Lett. A. {\bf 87}, 89 (1981), S. R. Maiti, Phys. Rev. D. {\bf 25}, 2518 (1982), B. Modak, J. Astrophys. Astron. {\bf 5}, 317 (1984), A. Banerjee, S. Choudhury and B. Bhui, Phy. Rev. D. {\bf 40}, 670 (1989), D. Schafer and H. F. Goenner, Gen. Relativ. Grav. {\bf 42}, 2119 (2000), S. Chakrabarti and N. Banerjee, Phys. Rev. D. {\bf 95}, 024015 (2017).

\bibitem{israel} W. Israel, Nuovo Cimento, {\bf 44B}, 1 (1966) ; Nuovo Cimento, {\bf 48B}, 463 (1966).

\bibitem{santosmatching} N. O. Santos, Mon. Not. R. Astron. Soc. {\bf 216}, 403 (1985).

\bibitem{chanmatching} R. Chan, Mon. Not. R. Astron. Soc. {\bf 316}, 588 (2000).

\bibitem{MS} C. W. Misner and D. Sharp, Phys. Rev. {\bf 136}, B571 (1964) ; M. Cahill and G. McVittie, J. Math. Phys. {\bf 11}, 1382 (1970).

\bibitem{cahill} M. E. Cahill and G. C. McVittie, J. Math. Phys., {\bf 11}, 1382 (1970).

\bibitem{simpson2} A. Simpson, P. Martin-Moruno and M. Visser, Class. Quant. Gravit., {\bf 36}, 145007 (2019). 

\bibitem{vaidyanewref} V. Husain, Phys. Rev. D. {\bf 53}, R1759 (1996) ; J. D. Brown and V. Husain, Int. J. Mod. Phys. D. {\bf 6}, 563 (1997) ; C. Barrabes and W. Israel, Phys. Rev. D. {\bf 43}, 1129 (1991).

\bibitem{wangvaidya} A. Wang and Y. Wu, Gen. Relativ. Grav. {\bf 31} (1), 107 (1999).

\bibitem{hawkingellis} S. W. Hawking and G. F. R. Ellis, The Large Scale Structure of Spacetime, (Cambridge University Press, Cambridge, 1973).

\bibitem{sk} S. Kar, Phys. Rev. {\bf D 49}, 862 (1994); S. Kar and D. Sahdev, Phys. Rev. {\bf D 53}, 722 (1996).
\end{thebibliography}
\end{document}